\documentclass[a4paper,fleqn,usenatbib]{mnras}

\usepackage{newtxtext,newtxmath}

\usepackage[T1]{fontenc}
\usepackage{ae,aecompl}

\usepackage{graphicx}	
\usepackage{amsmath}	
\usepackage{amssymb}	
\usepackage{nicefrac}
\usepackage{subfigure}
\usepackage{siunitx}
\usepackage{url}
\usepackage{enumerate}

\newcommand\Tstrut{\rule{0pt}{2.6ex}}       
\newcommand\Bstrut{\rule[-1.1ex]{0pt}{0pt}}
\newcommand{\appropto}{\mathrel{\vcenter{
  \offinterlineskip\halign{\hfil$##$\cr
    \propto\cr\noalign{\kern2pt}\sim\cr\noalign{\kern-2pt}}}}}
\defcitealias{G&T09}{G\&T09} 
\defcitealias{DL90}{D\&L90}
\defcitealias{Ghisellini10}{G10}   

\title[The jet of the $\gamma$-NLS1 1H~0323+342]{The relativistic jet of the $\gamma$-ray emitting narrow-line Seyfert 1 galaxy 1H~0323+342}

\author[D. Kynoch et al.]{Daniel Kynoch$^{1}$\thanks{E-mail: daniel.kynoch@durham.ac.uk}
Hermine Landt$^{1}$,
Martin J. Ward$^{1}$,
Chris Done$^{1}$,
Emma Gardner$^{1}$,
\newauthor
Catherine Boisson$^{2}$,
Maialen Arrieta-Lobo$^{2}$,
Andreas Zech$^{2}$,
Katrien Steenbrugge${^3}$
\newauthor
and Miguel Pereira Santaella$^{4}$
\\
$^{1}$Centre for Extragalactic Astronomy, Department of Physics, Durham University, South Road, Durham, DH1 3LE, UK\\
$^{2}$LUTH, Observatoire de Paris, CNRS, Universit\'{e} Paris Diderot, PSL Research University Paris, 5 place Jules Janssen, \\92195 Meudon, France\\
$^{3}$Instituto de Astronom\'{i}a, Universidad Cat\'{o}lica del Norte, Avenida Angamos 0610, Antofagasta, Chile \\
$^{4}$Department of Physics, University of Oxford, Keble Road, Oxford, OX1 3RH
}

\date{Accepted 2017 December 04. Received 2017 December 04; in original form 2017 August 23}

\pubyear{2017}

\begin{document}
\label{firstpage}
\pagerange{\pageref{firstpage}--\pageref{lastpage}}
\maketitle

\begin{abstract}
The detection of several radio-loud narrow-line Seyfert 1 (NLS1) galaxies by the \textit{Fermi Gamma-Ray Space Telescope} hints at the existence of a rare, new class of $\gamma$-ray emitting active galactic nuclei with low black hole masses. Like flat spectrum radio quasars (FSRQs), their $\gamma$-ray emission is thought to be produced via the external Compton mechanism whereby relativistic jet electrons upscatter a photon field external to the jet, e.g.\ from the accretion disc, broad line region (BLR) and dusty torus, to higher energies. Here we study the origin of the $\gamma$-ray emission in the lowest-redshift candidate among the currently-known $\gamma$-ray emitting NLS1s, 1H~0323+342, and take a new approach. We observationally constrain the external photon field using quasi-simultaneous near-IR, optical and X-ray spectroscopy. Applying a one-zone leptonic jet model, we simulate the range of jet parameters for which this photon field, when Compton scattered to higher energies, can explain the $\gamma$-ray emission. We find that the site of the $\gamma$-ray emission lies well within the BLR and that the seed photons mainly originate from the accretion disc. The jet power that we determine, $1.0\times10^{45}$~erg~s$^{-1}$, is approximately half the accretion disc luminosity. We show that this object is not simply a low-mass FSRQ, its jet is intrinsically less powerful than predicted by scaling a typical FSRQ jet by black hole mass and accretion rate.  That $\gamma$-ray emitting NLS1s appear to host underpowered jets may go some way to explaining why so few have been detected to date.

\end{abstract}





\section{Introduction}
The detection of several narrow-line Seyfert 1 (NLS1) galaxies
by the \textit{Fermi Gamma-Ray Space Telescope} suggests the existence
of a rare, new class of $\gamma$-ray emitting active galactic nuclei
(AGN). These are similar to the standard blazars in that their \textit{Fermi} 
$\gamma$-ray emission is dominated by a relativistic jet aligned close to the line
of sight, but distinctly different in that this is powered by
accretion onto a black hole (BH) of much lower mass  (\citealt{Abdo09b}; \citealt{Foschini11}). 
The mechanisms by which such relativisitic jets are launched and accelerated remain
poorly understood. 
These $\gamma$-NLS1s can provide new insights on how these processes might scale with BH mass. 

\textit{Fermi}-detected blazars can be divided into two types: BL Lacertae objects
(BL Lacs) and flat spectrum radio quasars (FSRQs). These can be
distinguished by their broad band spectral energy distributions (SEDs).
The BL Lac SEDs show two broad humps of
emission that are roughly equal in luminosity. One hump arises from synchrotron processes (generally peaking in the radio/IR/optical)
and the other from the synchrotron self-Compton mechanism (generally peaking in the GeV range) from
the same population of highly relativisitic electrons (with Lorentz factors
of up to $\gamma\sim 10^{5-6}$) accelerated within the jet. 
The entire SED is dominated by this emission, boosted along the line of sight by
the relativisitic bulk Lorentz factor ($\Gamma_\mathrm{BLF}\sim 10-20$) of the jet. In contrast, the
FSRQs have GeV Compton humps that are considerably more
luminous than their synchrotron emission humps, and they additionally show a
clear accretion disc spectrum as a third hump in the region between the
two jet emission components, together with an associated broad line
region (BLR). These differences can be understood in the context of a change
in the nature of the accretion flow, with the BL Lacs having low
accretion rates so the accretion flow is in the hot, advection
dominated state with little intrinsic UV emission and hence a very weak or absent
BLR. These advection-dominated accretion flows (ADAFs) collapse into a standard disc at luminosities
above a few percent of the Eddington luminosity, so the higher luminosities FSRQ have
a UV bright disc which provides the ionisation to produce a BLR which in turn gives an additional
source of seed photons external to the jet for Comptonisation
(external Compton: EC), leading to the observed much brighter Compton
hump (\citealt{GMT09}). The BH mass can be derived from standard BLR
scaling relations for the FSRQ, and is always found to be very high, with
$M_\mathrm{BH}\gtrsim10^8$~M$_\odot$ (e.g.\ \citealt{G&T15}). Conversely, it is much
more difficult to constrain in BL Lacs as these have very weak (or no) lines, but
studies of the host galaxies conclude that these are powered by
similarly high mass BHs (e.g.\ \citealt{Plotkin11}; \citealt{Falomo03}).
Together the BL Lacs and FSRQs form a standard `blazar sequence'
of increasing accretion power onto the most massive BHs
(\citealt{Ghisellini17}; \citealt{Fossati98}).
In terms of AGN unification schemes it is insightful to investigate whether $\gamma$-NLS1s represent the low-mass, 
low-power tail of FSRQs in this sequence, or whether they constitute a 
genuinely new class of their own.

At larger inclination angles Doppler de-boosting means that the jet does not dominate 
the SED but these objects are still easily identified by their strong radio emission 
from both the jet core and large-scale radio lobes. 
This is often quantified as a radio-loudness
parameter, defined from a radio-to-optical flux ratio
$R=f_{5~\mathrm{GHz}}/f_{B~\mathrm{band}}$, with $R>10$ defining a radio-loud (RL) quasar.
Radio-quiet (RQ) quasars and the lower-power Seyfert AGN can exhibit radio
jet structures, but these are slow, and poorly-collimated compared
with blazar jets (e.g.\ \citealt{Middelberg04}).

Early studies (e.g.\ \citealt{Laor00}; \citealt{McLure01}) found no evidence
of a RL-AGN population with BH masses
$M_\mathrm{BH}\lesssim10^8$~M$_\odot$. High-mass BHs are 
almost exclusively found in elliptical galaxies with large bulges,
leading to ideas that there is something about the evolutionary
history of these systems which triggers jet production such as BH
spin (\citealt{BZ77}) or the history of concentration of
magentic flux (\citealt{Sikora13}), or both. 
However, this simple paradigm is now being challenged by the discovery of
lower-mass RL-AGN (\citealt{Ho02}; \citealt{Yuan08}) which are instead
hosted by spiral galaxies. A few of these have been detected by \textit{Fermi},
and appear to form a low-mass, low-power tail of the FSRQ population,
with a detectable disc component and BLR, together with dominant EC
emission. The BLR line velocity widths are fairly narrow, often with FWHM
below 2000~km~s$^{-1}$ which forms the (arbitrary) cutoff for an object 
designated as a NLS1 (\citealt{OP85}).  Such low
velocities of the BLR gas indicate a low-mass BH accreting at a
high Eddington fraction. 
The percentage of NLS1s that are RL ($\approx7\%$,
\citealt{Komossa06} and $\approx5\%$, \citealt{Rakshit17}) is low when compared
with the RL fraction of AGN generally ($\approx15\%$), but a few (currently ten) 
of the RL-NLS1s have now been detected by \textit{Fermi} as $\gamma$-NLS1s
(\citealt{Abdo09a}; \citealt{Abdo09b}; \citealt{Donato11}; \citealt{Calderone12}; \citealt{Yao15-PKSJ1222}; \citealt{DAmmando15}; \citealt{DAmmando16}), confirming the presence of powerful, relativistic jets in these sources.


Whilst jet emission processes are relatively well-understood, the mechanisms
by which jets are launched and powered are still areas of intense research.
\cite{Ghisellini14} found a clear correlation between jet powers and accretion disc 
luminosities, but with the jet power exceeding the disc luminosity typically by a factor of ten.
The jet launching mechanism must therefore be very efficient and in some way linked to the accretion flow.
The $\gamma$-NLS1s are ideal objects to investigate this disc-jet connection, since 
they are nearby ($z<1$), very high accretion rate objects with luminous discs 
and blazar-like jets.

Here, we present a detailed study of the nearest $\gamma$-NLS1, 1H~0323+342\footnote{The J2000 name of this source, J0324+3410, is used in some other papers.} (RA: 03~24~41.16, Dec: +34~10~45.8), at a
redshift of $z=0.0625$ (\citealt{Landt17}).
High-energy $\gamma$-ray emission has been associated with its radio 
counterpart with high significance, and was
first reported by \cite{Abdo09b}.
In this paper we assemble an unprecedentedly well-sampled
SED containing several relatively high S/N spectra as well as complementary photometry.   
SEDs for this object have previously been presented in e.g.\ \cite{Abdo09b}, \cite{Paliya14} and \cite{Yao15},
but here we include much more spectral and photometric data to
assemble a more detailed and quasi-simultaneous SED. 
Our new approach is to use this SED to self-consistently determine the seed photons
available for the EC component, so we are able to better incorporate
the differences between the BLR size scale between this and the more
massive FSRQs. 

This paper is organised as follows: in \S~\ref{sec:mwl} we present the
multiwavelength data set we have assembled for this source and in
\S~\ref{sec:xray} we provide a detailed analysis of the
\textit{XMM-Newton} X-ray spectrum.  We describe how we use these data
to determine the ambient photon field contributions from the accretion
disc, X-ray corona, BLR and torus in \S~\ref{sec:epf}.  In
\S~\ref{sec:jetmod} we use a jet emission code to compute the
radiative output resulting from the jet's interaction with this photon
field and attempt to recover the jet parameters which best describe
the broadband SED.  The discussion and conclusions are presented in
\S~\ref{sec:disc} and \S~\ref{sec:conc}.  Throughout this paper, we
assume a $\Lambda$CDM cosmology with $H_0=70$~km~s$^{-1}$~Mpc$^{-1}$,
$\Omega_\mathrm{m}=0.3$ and $\Omega_\Lambda=0.7$.  Therefore the
redshift $z=0.0625$ implies a luminosity distance of 280~Mpc and a
flux-to-luminosity conversion factor of $9.41\times10^{54}$~cm$^2$.

\section{The multiwavelength data set}
\label{sec:mwl}
\begin{figure*}
\centering
\includegraphics[width=2.1\columnwidth, keepaspectratio]{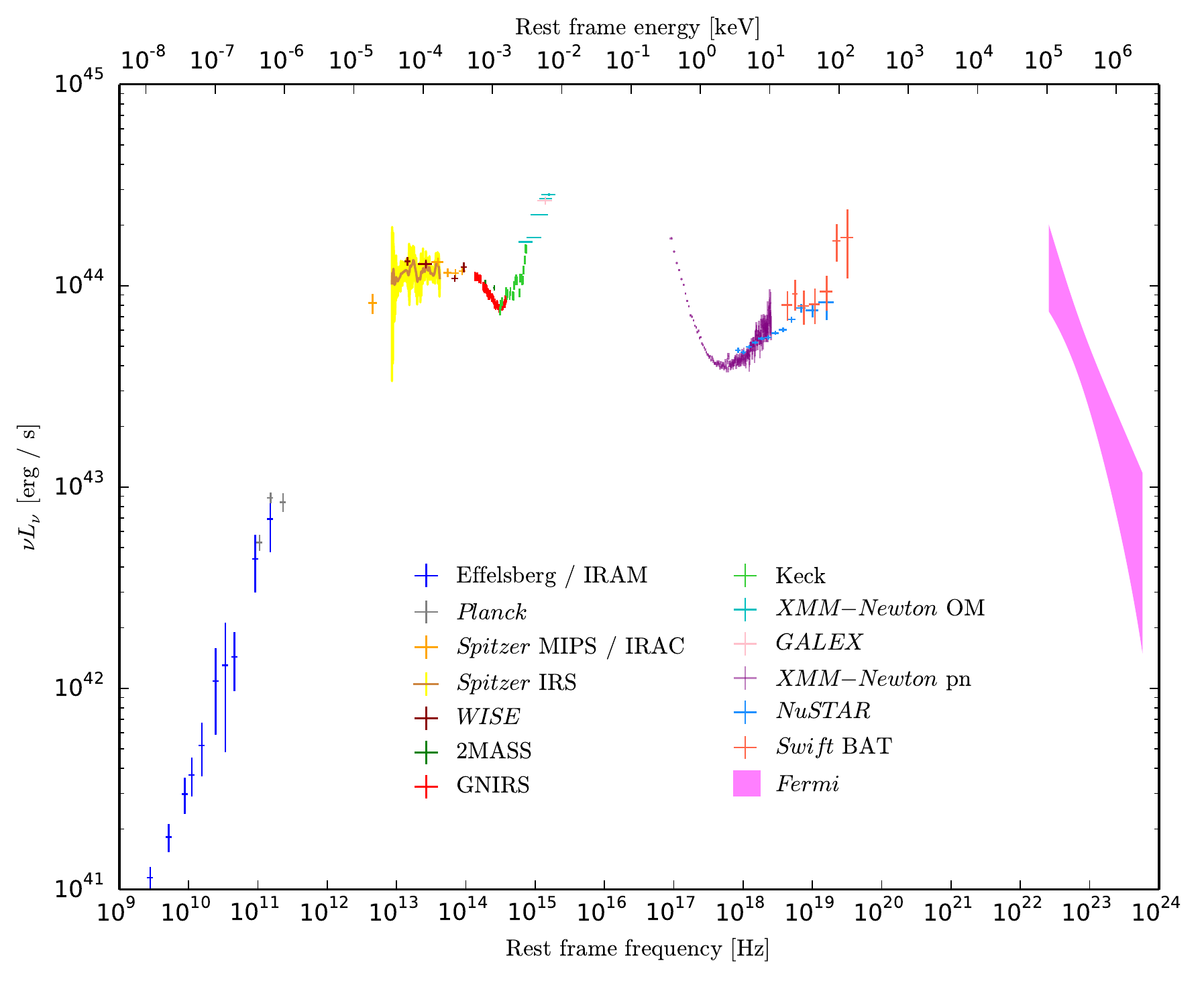}
\caption[General Caption]{The multiwavelength spectral energy distribution of the $\gamma$-NLS1 1H~0323+342.
The \textit{XMM-Newton} X-ray spectrum has been deabsorbed using a Galactic column $N_\mathrm{H}^\mathrm{Gal}=2.3\times10^{21}$~cm$^{-2}$.
The optical / UV data have been dereddened using an $A_V=0.71$.
For details of the original sources of these multiwavelength data, see Table~\ref{tab:data}.}
\label{fig:mwl}
\end{figure*}

Below we present the multiwavelength data set we have assembled for 1H~0323+342.
As a whole this data set is non-simultaneous, however parts of it are quasi-simultaneous.
In particular, the observations which we use to calculate the external photon field (including near infrared, optical and X-ray spectroscopy: see \S~\ref{sec:epf}) were all obtained in a time span of less than six months.     
The issue of variability is addressed in \S~\ref{sec:var} and \S~\ref{sec:dvar} and is the subject of a forthcoming paper (Arrieta-Lobo et al.\ 2017, in prep.). 
The data set spans an exceptionally wide range of frequencies, from $\sim10^9$~Hz in the radio up to $\sim10^{24}$~Hz in $\gamma$-rays. 
In addition the SED is also well-sampled, with data in the radio, infrared, optical, ultraviolet, X-rays and $\gamma$-rays. 
Because it is a bright source, much of these data are high S/N and includes spectra in the infrared, optical and X-ray as well as photometry.
Here, we present our new reductions / extractions of data from the \textit{Spitzer Space Telescope} (\S~\ref{sec:spitzer}); \textit{XMM-Newton} (\S~\ref{sec:xmm}) and \textit{Fermi} (\S~\ref{sec:Fermi}).
We also summarise the data which was used previously by \cite{Landt17}.
These data sets are supplemented by archival data from other facilities in the radio / sub-mm, infrared, ultraviolet and X-ray bands.
The complete data set and its reference sources are listed in Table~\ref{tab:data}.
The multiwavelength SED is shown in Fig.~\ref{fig:mwl}.   

\begin{table*}
	\centering
	\caption{The multiwavelength data set}
	\label{tab:data}
	\begin{tabular}{cllllllc}
	
Q	&	Band			& Instrument						& Observation date		& $\log(\nu_\mathrm{obs})$ & Flux 					& Luminosity			& Ref.\ \\
	&					& (Survey)							& (D/M/Y or M/Y)		& [Hz]				& [$10^{-14}$~erg/s/cm$^2$]		& [$10^{41}$~erg/s]		&  \\ 
\hline
			&	Radio			& Effelsberg						& 07/10--03/14			& $9.422$ 			& $1.22\pm0.16$					& $1.15\pm0.15$			& [1] \\
			&	Radio			& Effelsberg						& 07/10--03/14			& $9.686$			& $1.94\pm0.31$ 				& $1.82\pm0.29$			& [1] \\
			&	Radio			& Effelsberg						& 07/10--03/14			& $9.922$			& $3.17\pm0.64$					& $2.98\pm0.60$			& [1] \\
			&	Radio			& Effelsberg						& 07/10--03/14			& $10.02$			& $3.94\pm0.87$					& $3.71\pm0.82$			& [1] \\
			&	Radio			& Effelsberg						& 07/10--03/14			& $10.16$			& $5.5\pm1.7$					& $5.2\pm1.5$			& [1] \\
			&	Radio			& Effelsberg						& 07/10--03/14			& $10.36$			& $11.5\pm5.3$					& $10.8\pm5.0$			& [1] \\
			&	Radio			& Effelsberg						& 07/10--03/14			& $10.51$			& $13.8\pm8.7$					& $13.0\pm8.2$			& [1] \\
			&	Radio			& Effelsberg						& 07/10--03/14			& $10.63$			& $15.2\pm5.0$					& $14.3\pm4.7$			& [1] \\
			&	Radio			& IRAM								& 07/10--03/14		 	& $10.94$			& $47\pm1.5$					& $43.9\pm1.4$			& [1] \\
			&	Radio			& \textit{Planck}					& 08/09--11/10			& $11.00$ 			& $56.4\pm5.2$					& $53.1\pm4.9$			& [2] \\
			&	Radio			& IRAM								& 07/10--03/14		 	& $11.15$			& $73.7\pm2.3$					& $69.3\pm2.2$			& [1] \\
			&	Radio			& \textit{Planck}					& 08/09--11/10			& $11.16$			& $93.8\pm5.7$ 					& $88.2\pm5.4$			& [2] \\
			&	Radio			& \textit{Planck}					& 08/09--11/10			& $11.34$			& $89.3\pm9.4$					& $84.0\pm0.9$			& [2] \\
			&	Far-IR			& \textit{Spitzer} MIPS 			& 27/09/08				& $12.63$			& $869\pm8$						& $817\pm8$				& [3] \\
			&	Far-IR			& \textit{WISE} 					& 10--11/02/10			& $13.13$ 			& $1400\pm70$					& $1320\pm70$			& [4] \\
			&	Far-IR			& \textit{Spitzer} IRS$^\star$  	& 27/09/08				& $13.30$			& $1130\pm140$					& $1060\pm130$			& [3] \\
			&	Mid-IR			& \textit{WISE}  					& 10--11/02/10			& $13.41$ 			& $1360\pm70$					& $1280\pm70$ 			& [4] \\
			&	Mid-IR			& \textit{Spitzer} IRAC 			& 27/09/08				& $13.58$			& $1390\pm70$ 					& $1310\pm70$			& [3] \\
			&	Mid-IR			& \textit{Spitzer} IRAC 			& 27/09/08				& $13.72$			& $1230\pm60$ 					& $1160\pm60$			& [3] \\
			&	Mid-IR			& \textit{WISE}  					& 10/02/10--20/08/10	& $13.81$ 			& $1150\pm50$					& $1080\pm50$			& [4] \\
			&	Mid-IR			& \textit{Spitzer} IRAC 			& 27/09/08				& $13.82$			& $1230\pm60$ 					& $1150\pm60$			& [3] \\
			&	Mid-IR			& \textit{Spitzer} IRAC 			& 27/09/08				& $13.92$			& $1260\pm60$					& $1180\pm60$			& [3] \\
			&	Mid-IR			& \textit{WISE} 		 			& 10/02/10--20/08/10	& $13.95$ 			& $1310\pm80$					& $1230\pm80$			& [4] \\
			&	Near-IR			& (2MASS)		 		 			& 20/01/98				& $14.14$ 			& $1170\pm30$					& $1100\pm30$			& [5] \\
\checkmark	&	Near-IR			& GNIRS$^\star$					 	& 16/09/15				& $14.25$ 			& $1030\pm50$					& $970\pm50$			& [6] \\
			&	Near-IR			& (2MASS)		 		 			& 20/01/98				& $14.26$ 			& $1100\pm40$					& $1030\pm40$			& [5] \\
			&	Near-IR			& (2MASS)				 			& 20/01/98				& $14.39$ 			& $1040\pm30$					& $980\pm30$			& [5] \\
\checkmark	&	Optical			& Keck$^\star$						& 14/02/16				& $14.65$			& $950\pm50$					& $890\pm50$			& [6] \\
\checkmark	&	Optical			& \textit{XMM-Newton} OM			& 23/08/15				& $14.74$			& $1990\pm10$					& $1872\pm9$			& [3] \\
\checkmark	&	Optical			& \textit{XMM-Newton} OM			& 23/08/15				& $14.82$			& $1856\pm8$					& $1746\pm8$			& [3] \\
\checkmark	&	UV				& \textit{XMM-Newton} OM			& 23/08/15				& $14.94$ 			& $1963\pm8$					& $1847\pm8$			& [3] \\
\checkmark	&	UV				& \textit{XMM-Newton} OM			& 23/08/15				& $15.01$			& $2600\pm10$ 					& $2446\pm9$			& [3] \\
			&	UV				& \textit{GALEX} 				 	& 27/12/11	 			& $15.11$			& $2800\pm1000$   				& $2600\pm900$			& [7] \\
\checkmark	&	UV				& \textit{XMM-Newton} OM			& 23/08/15				& $15.11$			& $3010\pm30$ 					& $2830\pm30$			& [3] \\
\checkmark	&	UV				& \textit{XMM-Newton} OM			& 23/08/15				& $15.15$ 			& $3190\pm50$					& $3000\pm50$			& [3] \\
\checkmark	&	X-ray			& \textit{XMM-Newton} EPIC$^\star$	& 23/08/15				& $17.68$			& $430\pm10$					& $400\pm10$			& [3] \\
			&	X-ray			& \textit{NuSTAR}$^\star$ 		  	& 15--18/03/14			& $18.67$			& $720\pm20$					& $680\pm20$ 			& [6] \\
			&	X-ray			& \textit{Swift} BAT$^\star$		& 12/04--09/10			& $19.18$ 			& $995\pm200$					& $940\pm200$			& [8] \\
			&	$\gamma$-ray	& \textit{Fermi} LAT				& 01/08/15--30/09/15	& $22.39$			& $1500\pm450$ 					& $1400\pm420$ 			& [3] \\
			&	$\gamma$-ray	& \textit{Fermi} LAT				& 01/08/15--30/09/15	& $23.00$			& $370\pm130$					& $345\pm120$			& [3] \\
			&	$\gamma$-ray	& \textit{Fermi} LAT				& 01/08/15--30/09/15	& $23.74$			& $70\pm50$						& $65\pm50$	 			& [3] \\		
		\hline
	\end{tabular}
	\parbox[]{15.2cm}{\vspace{0.2em} $^\star$For spectra, we quote the flux at the indicated frequency at approximately the midpoint of each spectrum.
	The `Q' flag indicates the quasi-simultaneous data from which we determine the external seed photon field, as described in \S~\ref{sec:epf} in the text.  
	References: [1] \cite{Angelakis15}; [2] \textit{Planck} Second Point Source Catalog, \cite{PCCS2}; [3] this work; [4] \textit{WISE} AllWISE Source Catalog, \cite{WISE10}; [5] Two Micron All-Sky Survey, \cite{Skrutskie06}; [6] {\cite{Landt17}}; [7] \textit{GALEX} Data Release GR6, \citealt{Galex05}; [8] \textit{Swift} BAT 70-month All-Sky Hard X-ray Survey, \cite{Baumgartner13}.}
\end{table*}

\subsection{Quasi-simultaneous data sampling the external photon field}
Although not strictly simultaneous, we have obtained three spectra (in the infrared, optical and X-ray bands) and optical/UV photometry sampling the accretion flow which were taken over a period of less than six months.
In \S~\ref{sec:optxconv} we use these quasi-simultaneous data to parameterise the size scales and luminosities of the accretion disc, its X-ray corona and the hot dust emission from the torus, which (along with the BLR) contribute to the external photon field which is Compton upscattered by particles in the relativistic jet.  
In \S~\ref{sec:Fermi} we extract \textit{Fermi} $\gamma$-ray emission spanning a period from approximately a month either side of the \textit{XMM-Newton} observation.

\subsubsection{Gemini North}
The near-infrared spectrum, obtained in September 2015 using the Gemini Near-Infrared Spectrograph (GNIRS) on the Gemini North 8~m telescope, was presented in \cite{Landt17}. 
The average continuum S/N obtained in the $J$, $H$ and $K$ bands were $\sim40$, 70 and 90, respectively.
The spectrum was dereddened using the calculated extinction value $A_V=0.71$ from the Galactic neutral hydrogen column density $N_\mathrm{H}=1.46\times10^{21}$~cm$^{-2}$ given by \cite{DL90} (hereafter \citetalias{DL90}).

\subsubsection{Keck}
The optical spectrum was obtained using the Low Resolution Imaging Spectrometer mounted on the Keck 10~m telescope in February 2016.
As described in \cite{Landt17}, the average continuum S/N was $\sim60$ and we scaled up this spectrum in flux by $\approx40\%$ to match the near-IR spectrum.  

\subsubsection{\textit{XMM-Newton}}
\label{sec:xmm}
The large effective area of the \textit{XMM-Newton} X-ray observatory (\citealt{XMM}) makes it an excellent telescope with which to obtain high S/N X-ray spectra.
It carries three European Photon Imaging Camera (EPIC) detectors and a reflection grating spectrometer (RGS) which simultaneously conduct X-ray imaging and spectroscopy.
Its optical monitor (OM) operates concurrently with the X-ray detectors and can cycle through six filters covering optical/UV wavelengths.  
1H~0323+342 was observed by \textit{XMM-Newton} for 81~ks on 23-24 August 2015.
The three EPIC X-ray detectors (pn, MOS1 and MOS2) were operating in Large Window mode with the Medium filter in place.
Data from the observation (ID 0764670101; PI: D'Ammando) were obtained from the \textit{XMM-Newton} Science Archive
and the reduction was performed using the Science Analysis System (SAS, v15.0.0). 
 
We extracted \textit{XMM-Newton} OM photometry taken through all six filters using the SAS omichain and omsource tasks and standard procedures.
Fluxes were calculated from the count rates in each filter and dereddened using our derived $A_\mathrm{V}=0.71$ and adopting $R_\mathrm{V}=3.1$ and the reddening correction curves of \cite{Cardelli89}.
Data and response files for use in the spectral fitting package \textsc{Xspec} (\citealt{Arnaud96}) were generated using the flx2xsp tool.
Photometry from the V filter was excluded from our later analysis because it contains a strong emission line.
   
After filtering the EPIC event lists for flaring particle background we were left with good exposure times of 60, 72 and 70 ks for the pn, MOS1 and MOS2 detectors, respectively.     
Source spectra from all three detectors were extracted from a 20$^{\prime\prime}$-radius circular region centred on the source.
Background spectra were extracted from circular regions ($60^{\prime\prime}$ radius for pn and $40^{\prime\prime}$ for MOS) on an offset blank patch of sky on the same chip as the source.
Source count rates were 3.6, 0.94 and 1.0 counts~s$^{-1}$ in the pn, MOS1 and MOS2 detectors, respectively. 
The rate in pn exceeds the maximum rate of 1.5 counts~s$^{-1}$ for the avoidance of pile-up suggested in the Users Handbook.
A test for pile-up was performed using the SAS epatplot task and no evidence for pile-up was found.  
The extracted spectra were rebinned using the specgroup tool to achieve a minimum S/N of 5 in each channel and to not oversample the intrinsic instrumental energy resolution by a factor greater than 3. 
Because of the large number of counts in the spectra, this easily satisfied the requirement for a minimum of 20 counts per bin needed for $\chi^2$ analysis.
Our detailed X-ray spectral and temporal analyses are presented in \S~\ref{sec:xray}.

We also obtained the Pipeline Processing System (PPS) products from the two RGS instruments aboard \textit{XMM-Newton}.  
These instruments cover the 0.33--2.5~keV range at a much higher spectral resolution than the EPIC CCDs.
In our analysis we used only the first spectral orders (containing $\sim20000$ counts in total, see Table~\ref{tab:xmm-data}).

\begin{table}
	\centering
	\caption{Summary of \textit{XMM-Newton} exposures}
	\label{tab:xmm-data}
	\begin{tabular}{llll} 
	Detector	& Energy 	& Live time 	& Counts \\
				& [keV]		& [ks]			& \\
		\hline
	EPIC-pn		& 0.3--10	& 60			& 217650 \\
				& 0.33--2.5	&				& 182664 \\
	EPIC-MOS1	& 0.3--10	& 72			& 68017  \\
				& 0.33--2.5	&				& 54272 \\
	EPIC-MOS2 	& 0.3--10	& 70			& 73838	 \\
				& 0.33--2.5	&				& 58693 \\
	RGS1		& 0.33--2.5	& 80 			& 8912  \\
	RGS2		& 0.33--2.5	& 80 			& 10563 \\
		\hline
	\end{tabular}
\end{table}

\subsection{Additional data sampling the external photon field}
We supplement the data above with an infrared spectrum and photometry from \textit{Spitzer} which we attribute primarily to emission from the dusty torus (see \S~\ref{sec:spitzer} and \S~\ref{sec:disc}).
Additionally, we have photometry from \textit{WISE}, the 2MASS survey and \textit{GALEX} in the same frequency ranges as the \textit{Spitzer}, GNIRS and \textit{XMM-Newton} OM data, respectively.

\subsubsection{\textit{Spitzer}}
\label{sec:spitzer}
The \textit{Spitzer Space Telescope} (\citealt{Spitzer}) carries three scientific instruments.  
Its Infrared Array Camera (IRAC) images simultaneously at the wavelengths 3.6~$\mu$m, 4.5~$\mu$m, 5.8~$\mu$m and 8~$\mu$m; 
the Infrared Spectrograph (IRS) covers the wavelength range between $\approx5$--40~$\mu$m; 
the Multiband Imaging Photometer (MIPS) contains three arrays operating at 24~$\mu$m, 70~$\mu$m and 160~$\mu$m.

To obtain photometry from IRAC, we analysed the post-BCD (Basic Calibrated Data) images of 1H~0323+342 taken on the 27 September 2008 observation, available from the \textit{Spitzer} archive.
Using a 10$^{\prime\prime}$ aperture we determined fluxes at each of the four operating wavelengths with uncertainties $\approx5\%$.

\begin{figure}
    \includegraphics[width=\columnwidth, keepaspectratio]{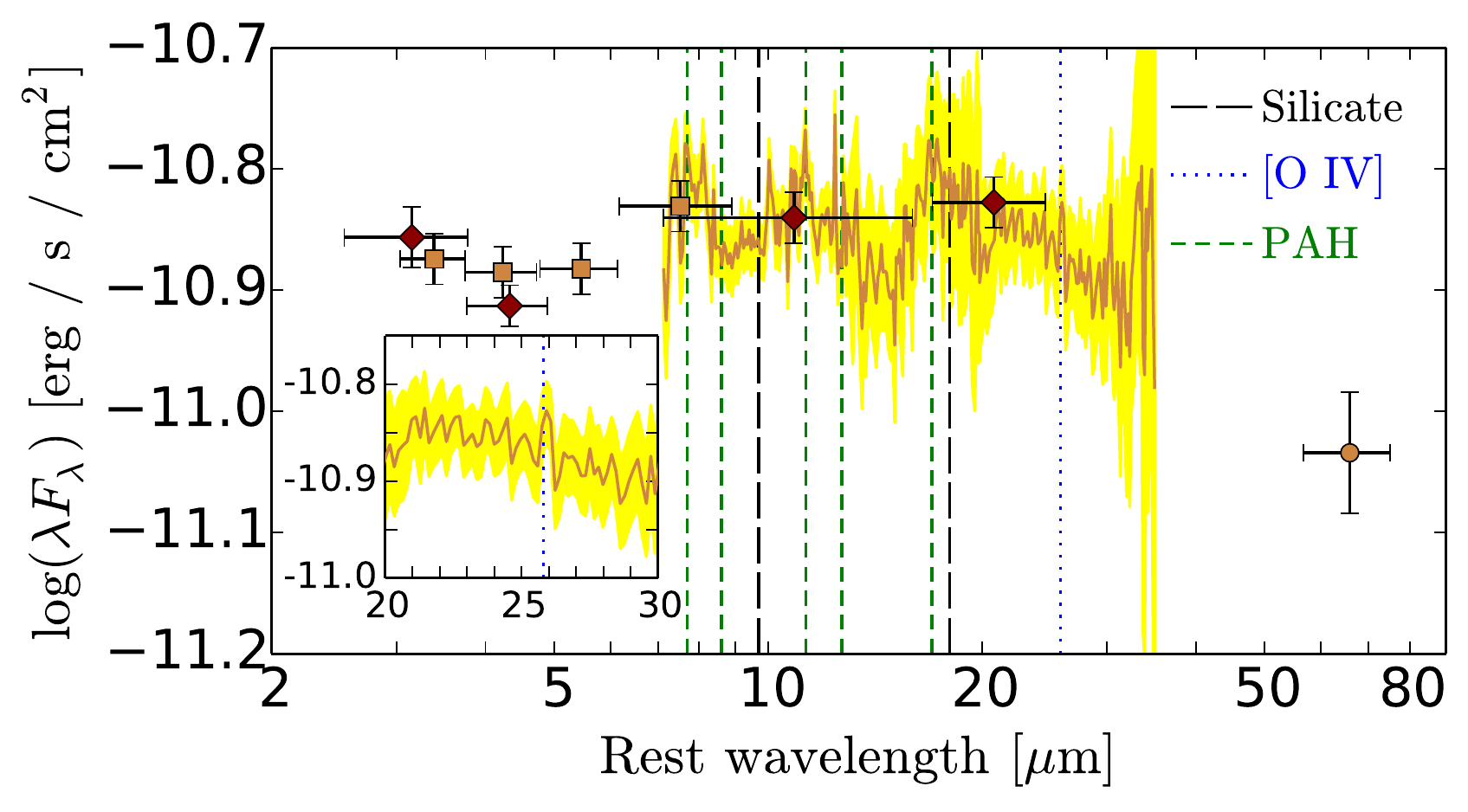}
\caption[General Caption]{\textit{Spitzer} spectroscopy and photometry of 1H~0323+342 taken in September 2008.  
The IRS spectrum is shown in ochre with its error region in yellow.  
Simultaneous photometry from IRAC and MIPS are shown with squares and a circle, respectively.
For comparison, later \textit{WISE} photometry points are shown with red diamonds.  
The wavelengths of the broad 9.7 and 18~$\mu$m silicate features are marked with black long dashed lines; the wavelengths of PAH features are marked with green short dashed lines and that of the [O~\textsc{iv}]~$\lambda25.8~\mu$m forbidden emission line is marked with a blue dotted line (this is also shown in the inset plot).}
\label{fig:Spitzer}
\end{figure}

The source was observed with IRS using the low spectral resolution (R$\sim$60--130) modules between 7.6 and 37.9 $\mu$m in the spectral mapping mode. 
We reduced this IRS mapping observation using the standard pipeline (version C18.18). 
First, we subtracted the background emission and removed rogue pixels using IRSCLEAN. 
Then, we projected the single IRS pointings into a grid similar to CUBISM (\citealt{Smith07}). 
From the data cube, we extracted the spectra using a $7.7^{\prime\prime}\times7.7^{\prime\prime}$ and a $17.8^{\prime\prime}\times17.8^{\prime\prime}$ square aperture centred at the nuclei in the short-low (SL; 7.6--14 $\mu$m) and long-low (LL; 14--36 $\mu$m) cubes, respectively. 
A point-source aperture correction was applied based on the IRS mapping observations of stars. There is a good agreement between the continuum levels at the overlapping spectral ranges of the different modules (SL and LL). 
This suggests, that the mid-IR emission is dominated by a point-like source at the spatial resolution of IRS ($\sim$2--9$^{\prime\prime}$, depending on the wavelength).  The IRS spectrum is shown in Fig.~\ref{fig:Spitzer} along with the MIPS and IRAC photometry.

We measured the possible [O~\textsc{iv}]~$\lambda25.89~\mu$m emission line in the IRS spectrum.
To do so, we fit the 20--32~$\mu$m region with a Gaussian profile and the underlying continuum as a power-law of the form $F_\lambda = a (\lambda / b)^{-c}$ where the constants $a$, $b$ and $c$ are free parameters in the fit.
We find the Gaussian line has a central, rest-frame wavelength of $25.92_{-0.04}^{+0.03}~\mu$m, consistent with the [O~\textsc{iv}] line.
From the fitted Gaussian, which has a $\mathrm{FWHM}=0.26\pm0.04~\mu\mathrm{m}$, we calculate an integrated luminosity in the line $\log(L_\mathrm{[O\textsc{iv}]})=41.3$~erg~s$^{-1}$.
 
1H 0323+342 was detected at 70~$\mu$m using MIPS.  
The data reduction was performed using the MOPEX analysis tool.  
Following the prescription in the MIPS instrument handbook v3.0, source counts were extracted from a circular region of radius 35$^{\prime\prime}$ and the background counts were taken from an annulus with inner and outer radii of 39$^{\prime\prime}$ and 65$^{\prime\prime}$, respectively.  
The photometric uncertainty was calculated using Eqn.~(1) in \cite{Carpenter08}.  
The source was found to have a flux density $207\pm2$~mJy, equivalent to a flux of $(8.69\pm0.08)\times10^{-12}$~erg~s$^{-1}$~cm$^{-2}$.
 
\subsubsection{WISE}
The \textit{Wide-field Infrared Survey Explorer} (\textit{WISE}; \citealt{WISE10}) telescope was launched in December 2009 with the aim of conducting an all-sky survey in the infrared.
It observes in four photometric bands simultaneously: W1 (3.4~$\mu$m), W2 (4.6~$\mu$m), W3 (12~$\mu$m) and W4 (22~$\mu$m).
Photometry for the source 1H~0323+342 was obtained in each of these bands from the AllWISE Source Catalog\footnote{\label{note:irsa}\url{http://irsa.ipac.caltech.edu/}}.
The photometric magnitudes were calculated from multiple observations (24 for W1 and W2, 12 for W3 and W4) recorded during the survey.
The observation start and end dates correspond to those listed in the online long form catalogue.

\subsubsection{2MASS}
1H~0323+342 was observed as part of the Two Micron All-Sky Survey (2MASS) which was conducted between 1997 and 2001.
We obtained measurements in the J, H and K$_\mathrm{S}$ bands from the 2MASS All-Sky Point Source Catalog\footnote{Also available from the Infrared Science Archive, see note \ref{note:irsa}.} (\citealt{Skrutskie06}).  
The S/N in these bands were 99.9, 95.3 and 131.0, respectively. 

\subsubsection{\textit{GALEX}}
The \textit{Galaxy Evolution Explorer} (\textit{GALEX}; \citealt{Galex05}) was a dedicated UV space telescope which launched in April 2003 and operated for ten years.
1H~0323+342 was detected in the near-UV by \textit{GALEX} during a 96~s exposure on 27 December 2011.  Data were extracted from the sixth \textit{GALEX} data release, GR6\footnote{\url{http://galex.stsci.edu/GR6/}}.  
The flux was dereddened using the same procedure as for the \textit{XMM-Newton} OM fluxes.

\subsection{Data sampling the jet emission}
At both the very low and very high frequency ends of the SED, we have data which sample the emission from the relatvistic jet.
\subsubsection{Effelsberg and IRAM}
Radio light curves and SEDs of 1H~0323+342 were produced as part of the \textit{Fermi}-GST Multiwavelength Monitoring Alliance (F-GAMMA, \citealt{F-GAMMA}) monitoring programme.  
The observations were conducted between 31 July 2010 and 11 March 2014.  
Flux densites at 2.64, 4.85, 8.35, 10.45, 14.60, 23.05, 32.00 and 43.05~GHz were obtained at the 100~m Effelsberg telescope.  
86.24 and 142.33~GHz readings were made at the 30~m IRAM telescope.  
For our SED, we have taken the mean flux densities and their standard deviations as reported in Table 8 of \cite{Angelakis15}\footnote{In this paper our source is named J0324+3410.}; we refer the reader to this paper for further details.

\subsubsection{\textit{Planck}}
We complemented the low-frequency data with non-simultaneous Planck observations taken from the latest version of the \textit{Planck} Catalog of Compact Sources\footnote{\url{http://irsa.ipac.caltech.edu/data/Planck/release_2/catalogs/}} (PCCS2, \citealt{PCCS2}) that compiles all sources, both Galactic and extragalactic, detected with high confidence over the full sky during the period between August 2009 and August 2013. This catalogue contains average intensity information for the sources which may have been observed more than once.

Using a cone search of $1^\prime$, clear association with 1H~0323+342 was found at 100 and 143~GHz in the good-quality PCCS2 catalogues, and in addition at 217~GHz, taken from the PCCS2E catalogue. The catalogue gives multiple flux density estimates, the source associated to  1H~0323+342 being clearly identified on the cutout images; the photometry reported in Table~\ref{tab:data} are from Gaussian fitting method.  

\subsubsection{\textit{Fermi}}
\label{sec:Fermi}
The Large Area Telescope (LAT, \citealt{Atwood09}) onboard the \textit{Fermi} satellite detects $\gamma$-ray photons with energies between 20 MeV and above 300 GeV. The source 1H~0323+342 is listed in the second catalogue of flaring $\gamma$-ray sources detected with the Fermi All-sky Variability Analysis\footnote{\url{https://fermi.gsfc.nasa.gov/ssc/data/access/lat/FAVA/}} (FAVA), a tool that blindly searches for transients over the entire sky observed by the LAT (\citealt{Abdollahi16}). We analysed a subset of those data over the period 1 August to 30 September 2015, covering the date of the \textit{XMM-Newton} observation, using the publicly available Science Tools v10r0p53 . It appears that the source was in a low state. 

Photons in a circular region of interest (RoI) of radius 10$^\circ$, centred on the position of 1H~0323+342, were considered. The PASS 8 instrument response functions (event class 128 and event type 3) corresponding to the P8R2\_SOURCE\_V6 response were used together with a zenith-angle cut of 90$^\circ$. The Galactic diffuse emission has been modelled using the file gll\_iem\_v06.fits (\citealt{Acero16}) and the isotropic background using iso\_P8R2\_SOURCE\_V6\_v06.txt. Assuming a power-law spectral shape for 1H~0323+342, a binned likelihood analysis yields a detection with a Test Statistic TS = 11.26 ($\approx3.4\sigma$) with a flux of $F_{0.1-100~\mathrm{GeV}} = (4.65\pm1.68)\times10^{-8}$~cm$^{-2}$~s$^{-1}$ and a photon index of $\Gamma= 2.98\pm0.33$. 

To do this the fit was performed iteratively, with all the sources from the 3FGL catalogue within 14$^\circ$ around 1H~0323+342 included, with fixed parameters for those more than 10$^\circ$ away to account for event leakage in the RoI due to the large PSF at low energies. In a second step, the sources contributing to less than a TS of 9 and to less than 5\% of the total number of counts in the RoI have their parameters frozen. The only free parameters in the end are those of sources less than 3$^\circ$ away from 1H~0323+342, if not frozen in the previous step and the normalisations of the Galactic and isotropic diffuse emissions.

\subsection{Supplementary X-ray data}
We present \textit{Swift} X-ray telescope (XRT) monitoring data covering the same frequency and time period as the \textit{XMM-Newton} observation. 
Finally, hard X-ray spectra from both \textit{NuSTAR} and \textit{Swift} burst alert telescope (BAT) bridge the frequency range between the \textit{XMM-Newton} X-ray spectra and the $\gamma$-ray emission recorded by \textit{Fermi}.

\subsubsection{\textit{Swift} XRT}
\textit{Swift} monitoring of the source was conducted from 2 August to 24 December 2015, with snapshot observations of approximately 2~ks durations taken with an average $\approx6$~day cadence.
We reduced the data from the twelve observations taken between 2 August to 29 September 2015, around the time of the 81~ks \textit{XMM-Newton} observation (see \S~\ref{sec:xmm}) and covering the period of the \textit{Fermi} observations we use in this paper (see \S~\ref{sec:Fermi}).
The XRT products were created using xrtpipeline v0.13.2.  
The source extraction regions were a $47^{\prime\prime}$-radius circle centred on the source (corresponding to the 90\% encircled energy radius at 1.5~keV) and the background regions were $141^{\prime\prime}$ circular regions offset from the source, in an area free of field sources.
The spectra were extracted using xselect and ancilliary response files were created with xrtmkarf.
The observations of 11 August and 29--30 September (OBS IDs 00036533056 and 00036533066) both had count rates slightly exceeding 0.5~counts~s$^{-1}$ and were investigated for pile up.
The wings of the PSF beyond $15^{\prime\prime}$ from the centre were fitted with a King function with the parameters $r_\mathrm{c}=5.8$ and $\beta=1.55$ fixed (see \citealt{Moretti05} for further details).
This function was then extrapolated into the inner regions.  
The deviation of the data from the model King function in the centre of the PSF was very marginal, so for our purposes it was unnecessary to extract the spectra from an annular region.
Using grppha, we rebinned each spectrum to contain a minimum of 20 counts per bin such that they were suitable for a $\chi^2$ analysis.
 
\subsubsection{\textit{NuSTAR}}
\label{sec:nustar}
A 200~ks exposure of the source was taken using \textit{NuSTAR} in March 2014.
The data reduction is detailed in \cite{Landt17}.
Here we use the co-added, time-averaged spectra from both focal plane modules FPMA and FPMB.     

\subsubsection{\textit{Swift} BAT}
We include catalogue data from the \textit{Swift} BAT seventy-month all-sky survey.  
The survey includes all sources detected in the hard X-ray energy range $14-195$~keV in the period December 2004 and September 2010 (\citealt{Baumgartner13}).  
The 14--195~keV photon index and flux were reported to be $\Gamma=1.73$ and $2.993\times10^{-11}$~erg~s$^{-1}$~cm$^{-2}$, respectively.

\section{X-ray analysis}
\label{sec:xray}

\subsection{Variability}
\label{sec:var}
\subsubsection{Short-term variability}
\begin{figure}
    \includegraphics[width=\columnwidth, keepaspectratio]{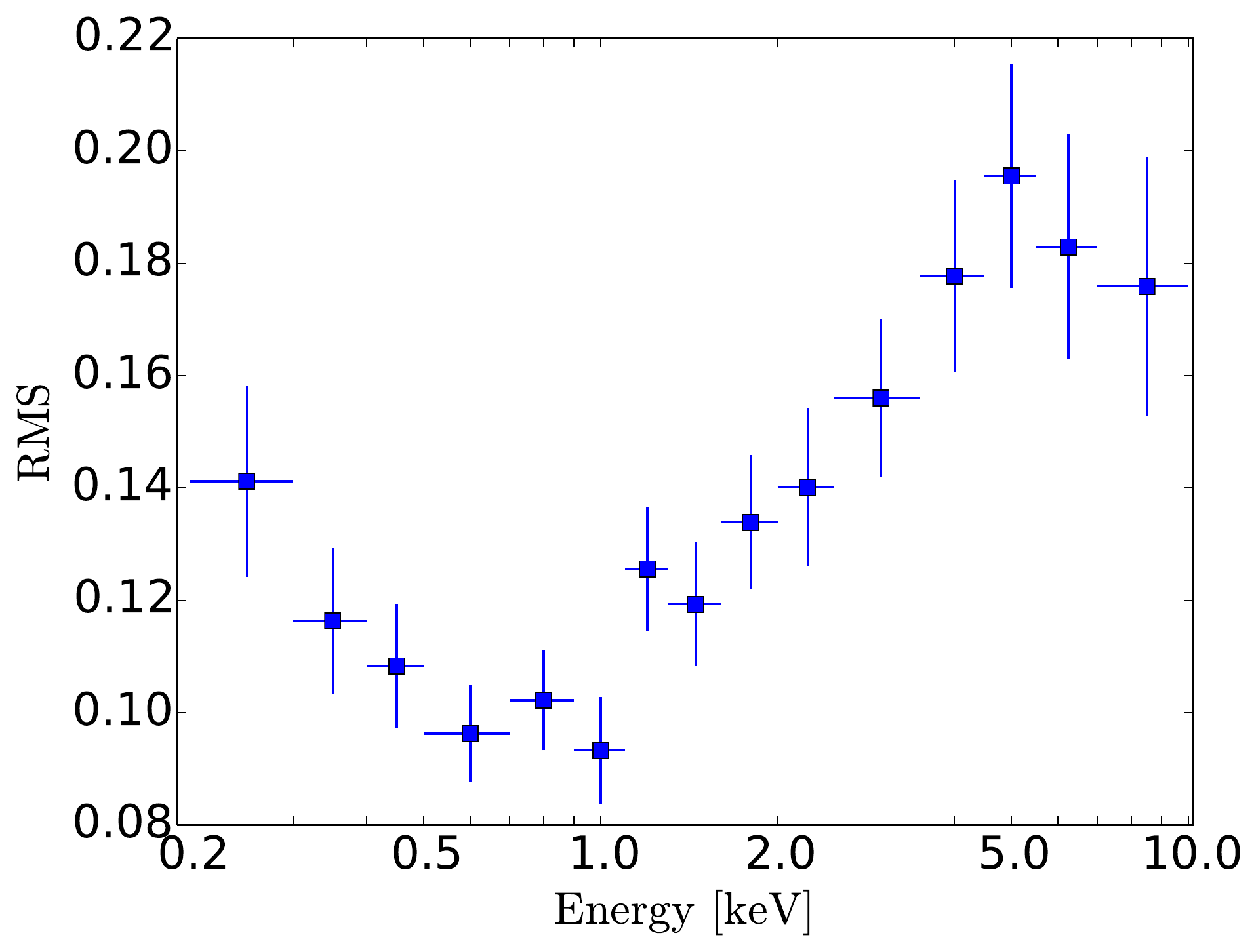}
\caption[General Caption]{The X-ray RMS fractional variability spectrum of 1H~0323+342.  
Fifteen lightcurves spanning the 0.2--10.0~keV energy range were created from the \textit{XMM-Newton} EPIC pn event file with time bins of 500~s.}
\label{fig:rms}
\end{figure}

We produced an RMS spectrum by creating lightcurves with 500~s time bins in fifteen energy bands between 0.2 and 10.0~keV.  
The excess RMS variability and its error were calculated for each lightcurve using the HEASARC ftool lcstats; these are plotted in Fig.~\ref{fig:rms}.
The spectrum clearly shows a break around 1~keV with the soft and hard spectral components exhibiting different variability behaviour.

\subsubsection{Medium-term variability}
\label{sec:med-var}
\begin{figure}
    \includegraphics[width=\columnwidth, keepaspectratio]{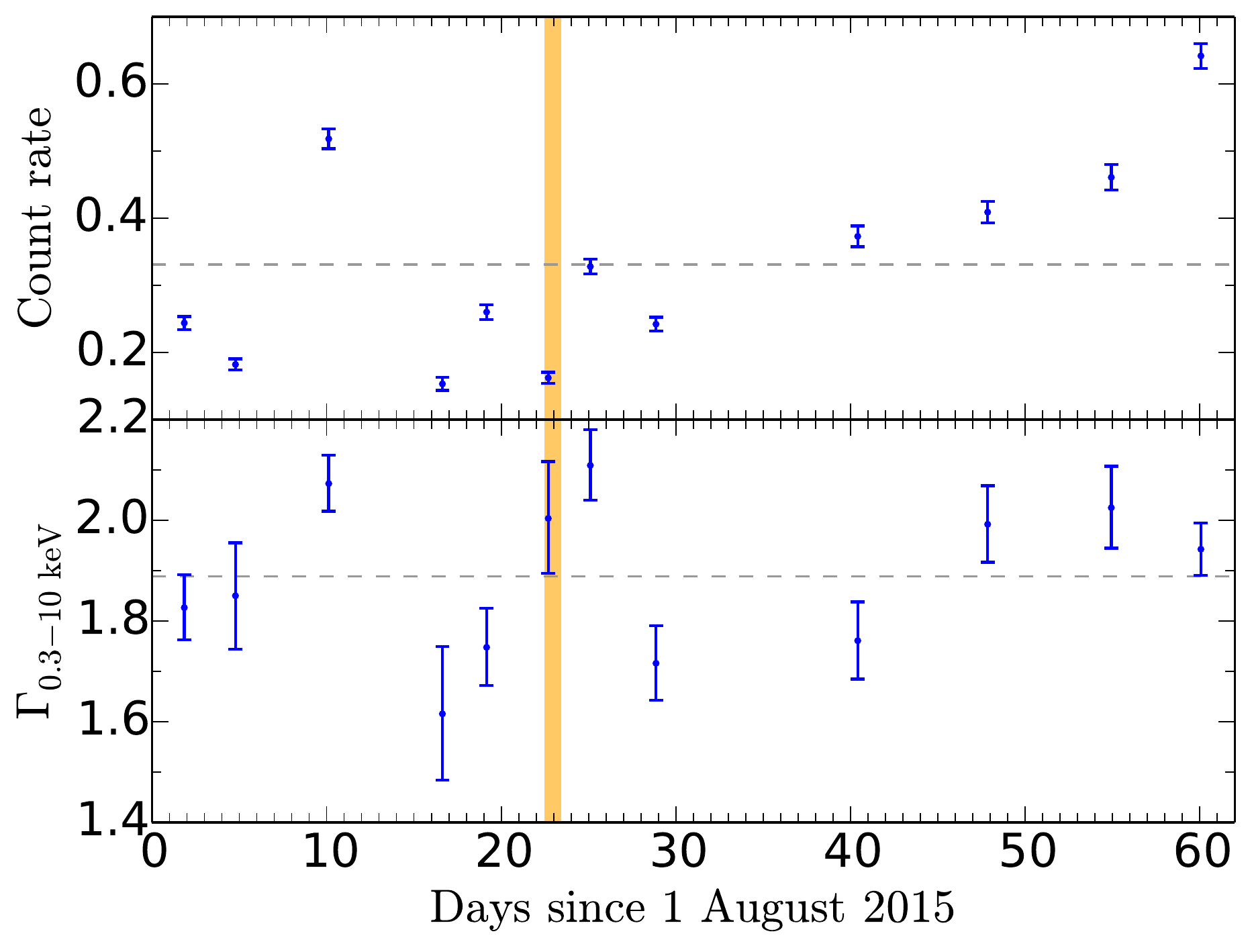}
\caption[General Caption]{Upper panel: \textit{Swift} XRT 0.3--10.0~keV count rates in counts~s$^{-1}$.  
Lower panel: The X-ray photon index $\Gamma$ of the best-fitting absorbed power-law model fit to the \textit{Swift} spectrum.
The mean values of both count rate and $\Gamma$ are shown as grey dashed lines.
The time span of the \textit{XMM-Newton} observation is highlighted in orange.}
\label{fig:Swift-lc}
\end{figure}

Each of the twelve \textit{Swift} spectra taken between 2 August and 30 September 2015 were fitted with a simple absorbed power-law model in \textsc{Xspec}.
To construct the lightcurve shown in Fig.~\ref{fig:Swift-lc} we report the 0.3--10.0~keV count rates and also the best-fit X-ray photon indices.
The count rates vary by a factor of four over this two-month period and the \textit{XMM-Newton} observation was taken during a period of particularly low activity.
The photon indices are poorly determined because of the limited S/N spectra, but by comparing the count rates and photon indices it can be seen that the source does not follow a simple `softer-when-brighter' pattern of behaviour.

\subsubsection{Longer-term variability}
The \textit{Swift} count rates in our main time interval of interest are $\approx0.3$~counts~s$^{-1}$ which are around the lowest values recorded in the five-and-a-half year lightcurve shown in \cite{Paliya14} (their Fig.~1). 
As noted in \cite{Landt17}, we observed only a $\sim30\%$ variation in $2-10$~keV flux between the three epochs of \textit{Swift} data (August 2013, December 2014 and September 2015) taken around the same time as our IR and optical spectra.  
In the corresponding \textit{Swift} UVOT data, some variability in the B, U, UVW1 and UVW2 filters was observed, but only $\sim20-30\%$ at the $2-3\sigma$ level.

\subsection{X-ray spectral analysis}
The X-ray spectral fitting of the \textit{XMM-Newton} data was performed in \textsc{Xspec} v12.9.0n (\citealt{Arnaud96}).
In all models we included a Galactic absorbing column (\textsc{phabs}), initially adopting the \citetalias{DL90} value $N_\mathrm{H}^\mathrm{Gal}=1.46\times10^{21}$~cm$^{-2}$.  
Cross-normalisation factors were included to account for differences in calibration between the three EPIC detectors; these did not vary by more than 5\%.

\subsubsection{The shape of the X-ray spectra}
\label{sec:xspec}
\begin{figure}
\begin{center}
	\includegraphics[width=\columnwidth, keepaspectratio]{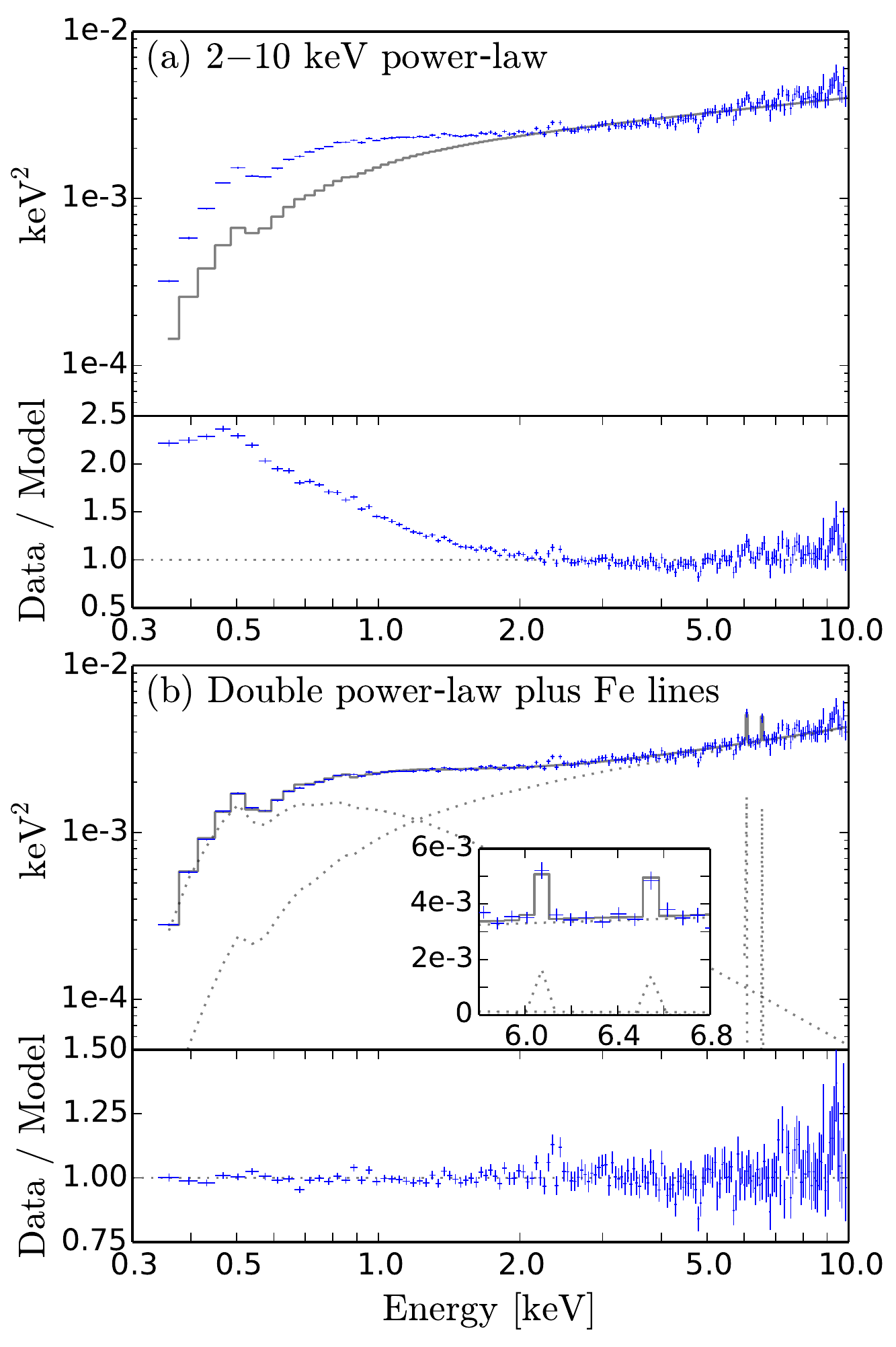}
\end{center}
\caption{Fits to the \textit{XMM-Newton} EPIC-pn X-ray spectrum of 1H~0323+342 taken during the 23--24 August 2015 observation.
Fits were performed to all three spectra (pn, MOS1 and MOS2) simultaneously; for clarity we show only the pn data here.
Upper panels show the data (crosses) with the total model (histograms) and the individual model components (dotted lines) for each of the three detectors in units of photons~cm$^{-2}$~s$^{-1}$~keV$^{-1}$; lower panels show the data / model ratios.
The model shown in (a) includes fixed $N_\mathrm{H}^\mathrm{Gal}=1.46\times10^{21}$~cm$^{-2}$; the model shown in (b) has free $N_\mathrm{H}^\mathrm{Gal}=2.33\times10^{21}$~cm$^{-2}$.}
\label{fig:xspec-new}
\end{figure}

A single power-law (with $\Gamma_{2-10~\mathrm{keV}}\approx1.7$) fit to the 2--10~keV data shows an excess of soft emission below $\approx2$~keV (see Fig.~\ref{fig:xspec-new}(a)).  
Consequently, a single power-law (with $\Gamma\approx2.1$) to the whole 0.3--10~keV range results in a very poor fit to the data with a reduced chi-squared $\chi^2_\nu=9.69$.
These fits, and the shape of the RMS spectrum shown in Fig~\ref{fig:rms}, clearly indicate that a continuum model with at least two components is required to fit the data.

A double power-law model (the first model in Table~\ref{tab:xmm}) is not a very good fit to the data.
In the course of our modelling, we noticed that our models overpredict the data at energies below $\approx0.5$~keV.
Additionally, the deabsorbed spectra do not rise towards lower energies to connect smoothly to the contemporaneous optical/UV photometry.
These issues could be resolved by including some additional absorption in our models.  
Allowing $N_\mathrm{H}^\mathrm{Gal}$ to be a free parameter we consistently find it rises to a value $\approx2.2\times10^{21}$~cm$^{-2}$, approximately 50\% greater than the \citetalias{DL90} value, and then gives a statistically significant improvement in the fits.
We note that these values are similar to the total (H~\textsc{i} plus H$_2$) Galactic column of $2.17\times10^{21}$~cm$^{-2}$ found by \cite{Willingale13}.
A double power-law model with free $N_\mathrm{H}^\mathrm{Gal}$ is a significant improvement with $\Delta\chi^2=162$ and an $F$-test probability $>99.99\%$.
The implications of deabsorbing the EPIC X-ray spectrum with this higher $N_\mathrm{H}^\mathrm{Gal}$ are discussed further in \S~\ref{sec:disc}.
The fit can be further improved by the inclusion of two narrow emission lines, as is described in the next section. 

\begin{table}
	\centering
	\caption{Results of X-ray spectral fits}
	\begin{tabular}{lll} 
		\hline
														
		Model 							& Parameter		 							& Value \\ 
		\hline
		\Tstrut\Bstrut

		\sc{phabs} $\times$				& $N_\mathrm{H}^\mathrm{Gal}$ [cm$^{-2}$] 	& $(1.46)\times10^{21}$~$^\mathrm{f}$ \\	
		(\sc{powerlaw} +				& $\Gamma_1$								& $2.54\pm0.02$ \\
										& norm.										& $\left(2.47_{-0.05}^{+0.04}\right)\times10^{-3}$ \\
		\sc{powerlaw})			 		& $\Gamma_2$			 					& $1.06\pm0.04$ \\
										& norm.										& $(4.7\pm0.4)\times10^{-4}$ \\
										& $\chi^2$/d.o.f. 							& 731/496 = 1.47 \vspace{0.5em}\\
		
		\sc{phabs} $\times$				& $N_\mathrm{H}^\mathrm{Gal}$ [cm$^{-2}$] 	& $(2.31\pm0.08)\times10^{21}$ \\	
		(\sc{powerlaw} +				& $\Gamma_1$								& $3.54\pm0.09$ \\
										& norm.										& $(2.10\pm0.04)\times10^{-3}$ \\
		\sc{powerlaw})			 		& $\Gamma_2$			 					& $1.49_{-0.03}^{+0.02}$ \\
										& norm.										& $(1.34\pm0.06)\times10^{-3}$ \\
										& $\chi^2$/d.o.f. 							& 569/495 = 1.15 \vspace{0.5em}\\
										
		\sc{phabs} $\times$				& $N_\mathrm{H}^\mathrm{Gal}$ [cm$^{-2}$] 	& $(2.33\pm0.08)\times10^{21}$ \\	
		(\sc{powerlaw} +				& $\Gamma_1$								& $3.59\pm0.09$ \\
										& norm.										& $(2.06\pm0.04)\times10^{-3}$ \\
		\sc{powerlaw} +			 		& $\Gamma_2$			 					& $1.52\pm0.02$ \\
										& norm.										& $\left(1.39_{-0.07}^{+0.06}\right)\times10^{}$ \\
		\sc{zgauss} +					& $E$ [keV]									& $6.43_{-0.02}^{+0.03}$ \\
		 								& norm.										& $(3.4\pm0.8)\times10^{-6}$ \\
										& EW [eV]									& $34\pm8$ \\
		 \sc{zgauss})					& $E$ [keV]									& $6.95\pm0.04$ \\
		 								& norm.										& $(2.4\pm0.8)\times10^{-6}$ \\
										& EW [eV]									& $28\pm9$ \\								
										& $\chi^2$/d.o.f. 							& 540/491 = 1.10 \vspace{0.5em}\\
					 			
		\hline
	\end{tabular}
\parbox[]{7cm}{\vspace{0.2em} $^\mathrm{f}$Parameter was frozen during the fitting procedure. 
Errors are quoted at the 1$\sigma$ level.  
The best-fit model is plotted in Fig.~\ref{fig:xspec-new}.}
\label{tab:xmm}
\end{table}

\subsubsection{Iron line emission features in the X-ray spectra}
\label{sec:FeK}
\begin{figure}
	\begin{center}
	\includegraphics[width=\columnwidth, keepaspectratio]{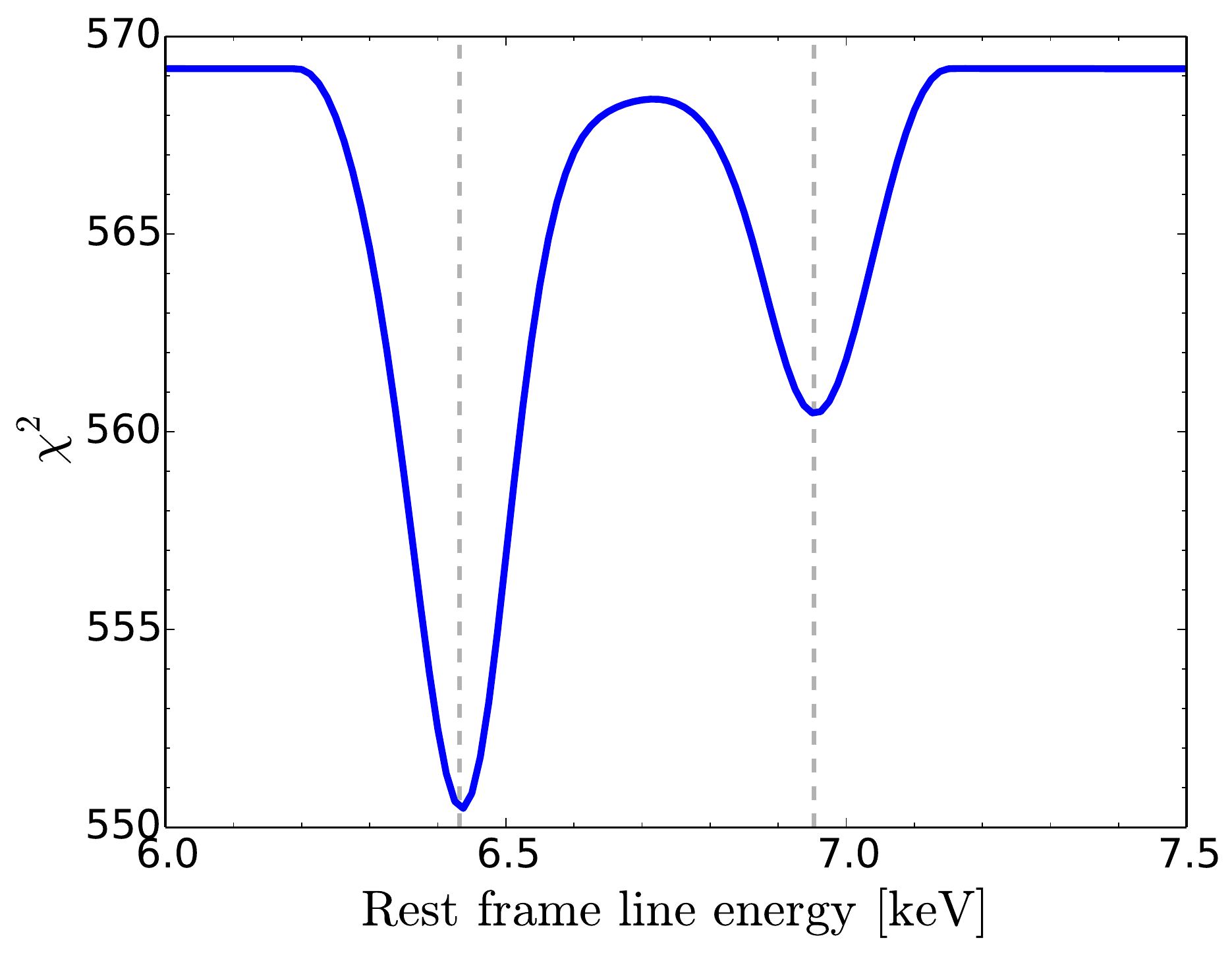}
	\caption{Variation in the $\chi^2$ fit statistic with the rest frame line energy of a narrow (fixed width $\sigma=10$~eV) Gaussian emission line.
	Best-fit line energies of 6.43 and 6.95~keV are indicated with dashed lines.}
	\label{fig:steppar}
	\end{center}
\end{figure}

Fig.~\ref{fig:steppar} shows that the fit statistic can be further improved by the addition of narrow emission lines at $\approx6.4$ and $\approx6.9$~keV.
We first added a broad line at $\approx6.4$~keV, but the fitting procedure reduced the width of the line to below the detector resolution, which is unphysical, and so instead we fit a narrow line of fixed width $\sigma=10$~eV.
We find that its rest-frame energy is $6.43_{-0.02}^{+0.03}$~keV, consistent with neutral Fe~K$\alpha$ emission, and inconsistent with the 6.7~keV energy of Fe~\textsc{xxv}.
The fit is improved by a $\Delta\chi^2=19$ for three additional free parameters to $\chi^2_\nu=550/493=1.12$, giving an $F$-test probability of 99.97\% compared to the model with no emission line.
The line flux is $\left(3.1_{-0.4}^{+0.7}\right)\times10^{-14}$~erg~s$^{-1}$~cm$^{-2}$ and its equivalent width (EW) is low at $34\pm8$~eV, which we discuss in \S~\ref{sec:disc}.

The fit is improved by a further $\Delta\chi^2=10$ with the inclusion of a second narrow Gaussian at $6.95\pm0.04$~keV, consistent with Fe~\textsc{xxvi}.
Clearly this is a weaker line than the neutral Fe~K$\alpha$ and we estimate its EW to be $28\pm9$~eV.
Our final X-ray spectral fit has a $\chi^2_\nu=1.10$, its parameters are given in Table~\ref{tab:xmm} and it is shown in Fig.~\ref{fig:xspec-new}(b).

\section{The origin of the $\gamma$-ray emission}
The $\gamma$-ray emission from high accretion-rate blazars such as FSRQs and $\gamma$-NLS1s is thought to be produced by the external Compton (EC) mechanism whereby an ambient field of soft seed photons external to the jet is Compton upscattered by relativistic leptons within the jet.
Emission from the accretion disc and its X-ray corona, the BLR and dusty torus can all potentially contribute to this external seed photon field.
Our new approach here is to determine the external photon field from our quasi-simultaneous IR-to-X-ray data which also samples the accretion flow.
Our parameterisation of the external photon field is presented below in \S~\ref{sec:epf}.
In \S~\ref{sec:jetmod} we then use a jet emission code to upscatter the external photon field and fit this to the full multiwavelength SED, determining the site of the $\gamma$-ray emission and the dominant source of seed photons.

\subsection{Determining the external photon field}
\label{sec:epf}
It is common in modelling EC emission to assume a standard external seed photon field which is upscattered by particles in the relativistic jet. 
Instead, we determine the external photon field of this particular source from our data.
In Table~\ref{tab:scale-epf} we summarise our findings and compare these to the standard assumptions made in the modelling of the photon field by \cite{G&T09} (hereafter \citetalias{G&T09}), on which our jet emission code is based.

\subsubsection{The accretion flow emission}
\label{sec:optxconv}
The accretion flow emission is dominated by radiation from a disc of material accreting onto the BH. 
This emission results from the radiative release of gravitational potential energy via viscous forces in the disc. 
The radiative efficiency $\eta$ of the accretion disc is determined by the location of its innermost stable circular orbit $R_\mathrm{isco}$, inside of which material plunges into the BH.
For a given BH mass and emissivity profile, discs with a smaller $R_\mathrm{isco}$ have a greater radiating surface area and hence a greater $\eta$. 
The location of $R_\mathrm{isco}$ is set by the spin\footnote{Here we use the dimensionless spin parameter $a_\star=Jc/GM_\mathrm{BH}^2$ where $J$ is the angular momentum of the BH.} of the BH; 
a maximally-spinning BH has $R_\mathrm{isco}=1R_\mathrm{g}=GM_\mathrm{BH}/c^2$ a non-rotating BH has $R_\mathrm{isco}=6R_\mathrm{g}$.

For low-mass and high accretion-rate BHs, the Wien tail of the accretion disc emission can extend into the soft X-ray bandpass.
However, the accretion discs of supermassive BHs are not generally expected to emit much X-radiation.
Most of the observed X-ray emission results from the Compton upscattering of photons by populations of hot electrons near to the BH.
One such region is the optically thin corona of the accretion disc, which produces X-ray emission well-represented as a power-law extending up to $\approx150$~keV.
As well as this power-law, many AGN also show evidence of a second Comptonisation region which is cooler and optically thicker than the corona.
The emission from this region is observed as an excess of soft X-ray emission above the coronal power-law, so it is often dubbed the `soft excess'.
When modelled as a thermal component, the soft excess has a remarkably constant temperature (0.1--0.2~keV) across sources covering a wide range of BH masses and Eddington ratios (e.g.\ \citealt{GierlinskiDone04}; \citealt{Porquet04}).

The continuum emission from the accretion flow of many AGN can therefore be represented by three components: the accretion disc emitting mostly in the optical/UV, plus a two Comptonisation regions producing soft excess and coronal X-rays.
An energy-conserving version of this simple concept (\textsc{optxagnf}, included in the current version of \textsc{Xspec}) is described by \cite{Done12}.
It includes a number of modifications to the simpler \cite{SS73} accretion disc spectrum which are relevant to the modelling of the accretion flows of NLS1s.
Firstly, for low-mass, high accretion rate systems such as NLS1s the inner disc is very hot and is not fully thermalised at all radii.  
The code applies an appropriate colour-temperature correction to the accretion disc spectrum.    
Secondly, the standard disc does not extend all the way down to $R_\mathrm{isco}$; instead it truncates at the coronal radius $R_\mathrm{cor}$.  Inside of $R_\mathrm{cor}$ a fraction $f_\mathrm{pl}$ of the power emerges as the coronal power-law emission.  
The remaining fraction $(1 - f_\mathrm{pl})$ of the power produces the soft excess.
Here we use the \textsc{Xspec} local model \textsc{optxconv} (\citealt{Done13}), an extension of \textsc{optxagnf} that approximates relativistic corrections to the spectrum, which are particularly pronounced at low inclinations and high spins.

At wavelengths longer than 1~$\mu$m, the Wien tail of blackbody emission from hot dust in the torus is dominant over the accretion disc emission.
The 1~$\mu$m region is covered by both our GNIRS (near-infrared) and Keck (optical) spectra; 
we extracted from these spectra data points sampling the emission line free continuum so that we can also parameterise the hot dust emission.

We include our \textit{NuSTAR} spectrum, taken 17 months prior to the \textit{XMM-Newton} observation, thereby extending our SED up to 79~keV.
We note that although similar in levels of flux, the photon index of our \textit{NuSTAR} spectrum ($\Gamma=1.80\pm0.01$) is softer than the index we determine in the overlapping energy range of \textit{XMM-Newton} spectrum ($\Gamma_{3-10~\mathrm{keV}}=1.59\pm0.02$).
It is known that a calibration issue with \textit{XMM-Newton} results in harder spectral indices above $\approx3$~keV than those determined from other X-ray telescopes.
For example, \cite{Ingram17} found the spectral index of their \textit{XMM-Newton} spectrum was $\Delta\Gamma=0.22$ lower than that of their \textit{NuSTAR} spectrum taken simultaneously, very similar to the discrepancy we see here.
In our non-simultaneous data the difference in spectral shape could be due to this miscalibration, but may of course result from a genuine spectral evolution between the two observations.

The mass accretion rate $\dot{M}$ through the outer accretion disc is constrained by the observed optical continuum emission.
We set the outer accretion disc radius to be equal to the self-gravity radius $R_\mathrm{sg}$, beyond which the disc fragments.
The X-rays are emitted from a region between $R_\mathrm{cor}$ (a model parameter which we fit) and $R_\mathrm{isco}$, the latter being determined by $a_\star$.
Since we have no prior input on $a_\star$ (from e.g.\ broad Fe~K$\alpha$) we test both zero- and high-spin cases with $a_\star$ fixed to 0.0 or 0.8.
As well as fitting a model in which all of the hard X-ray emission originates from the corona, we also fit models which include a hard X-ray contrbiution from the jet.
We model the jet as a broken power-law to allow for some curvature in its shape over the broad energy range.
In all models we fix $f_\mathrm{pl}$ to 0.3 (\citealt{Done12}; \citealt{Jin12-I}).
From our models we are able to determine several parameters which we will use to set the external photon field; namely: the size scales and luminosities of the accretion disc, its corona, and the hot torus dust, as well as the temperature of the dust (see \S~\ref{sec:tor}). 

\begin{figure}
\begin{center}
	\begin{tabular}{c}
	\includegraphics[width=\columnwidth, keepaspectratio]{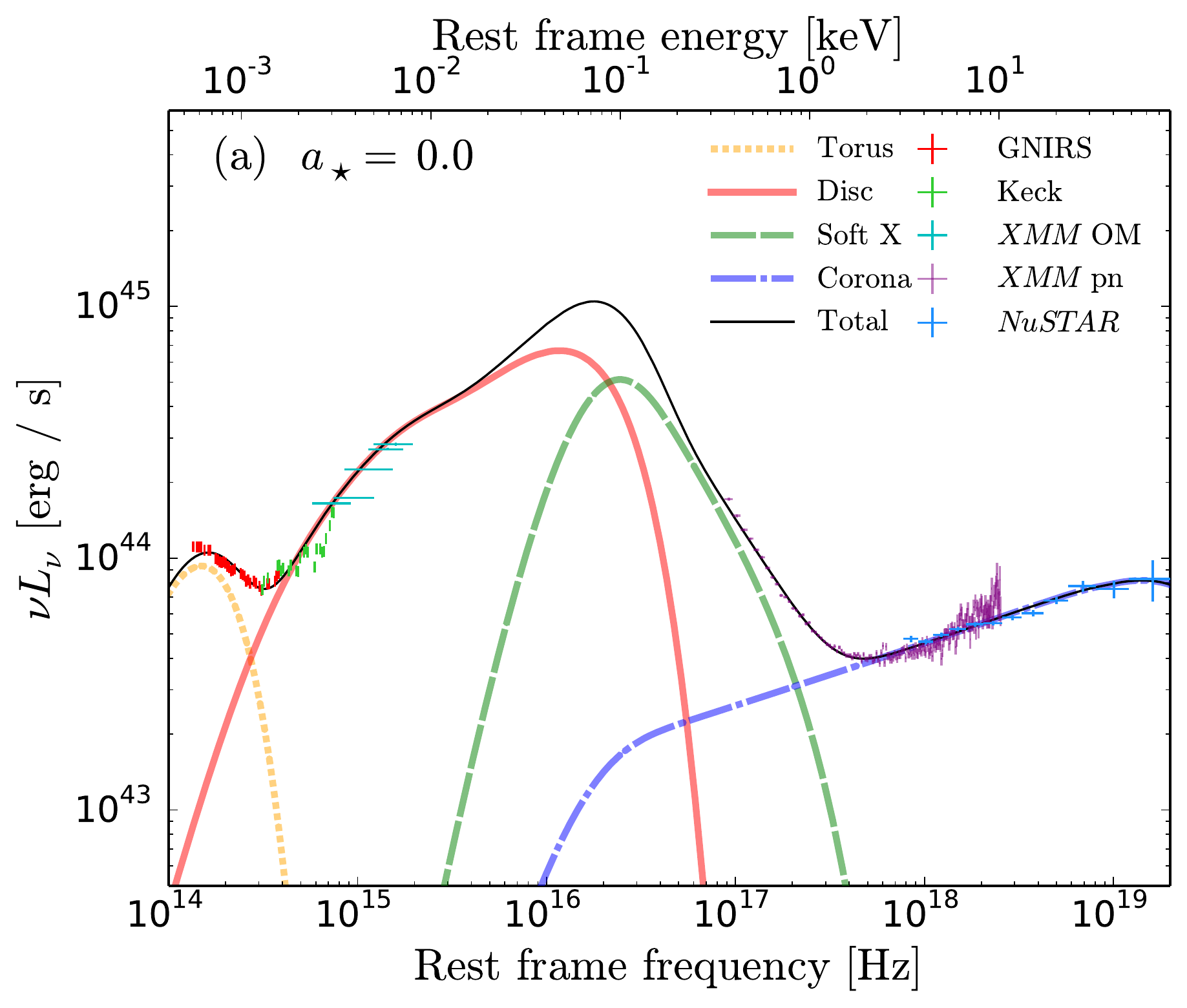} \\
	\includegraphics[width=\columnwidth, keepaspectratio]{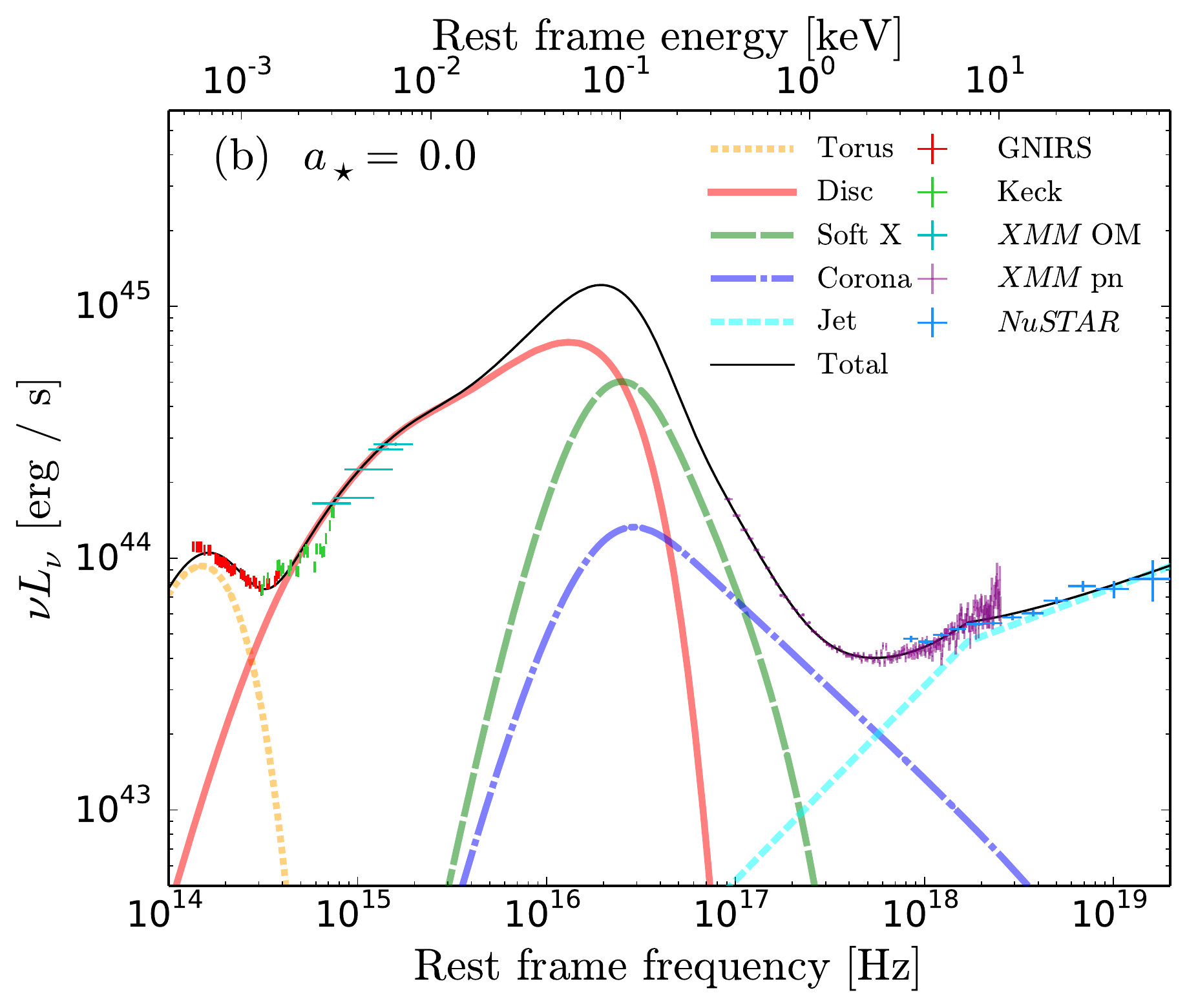} \\ 
	\includegraphics[width=\columnwidth, keepaspectratio]{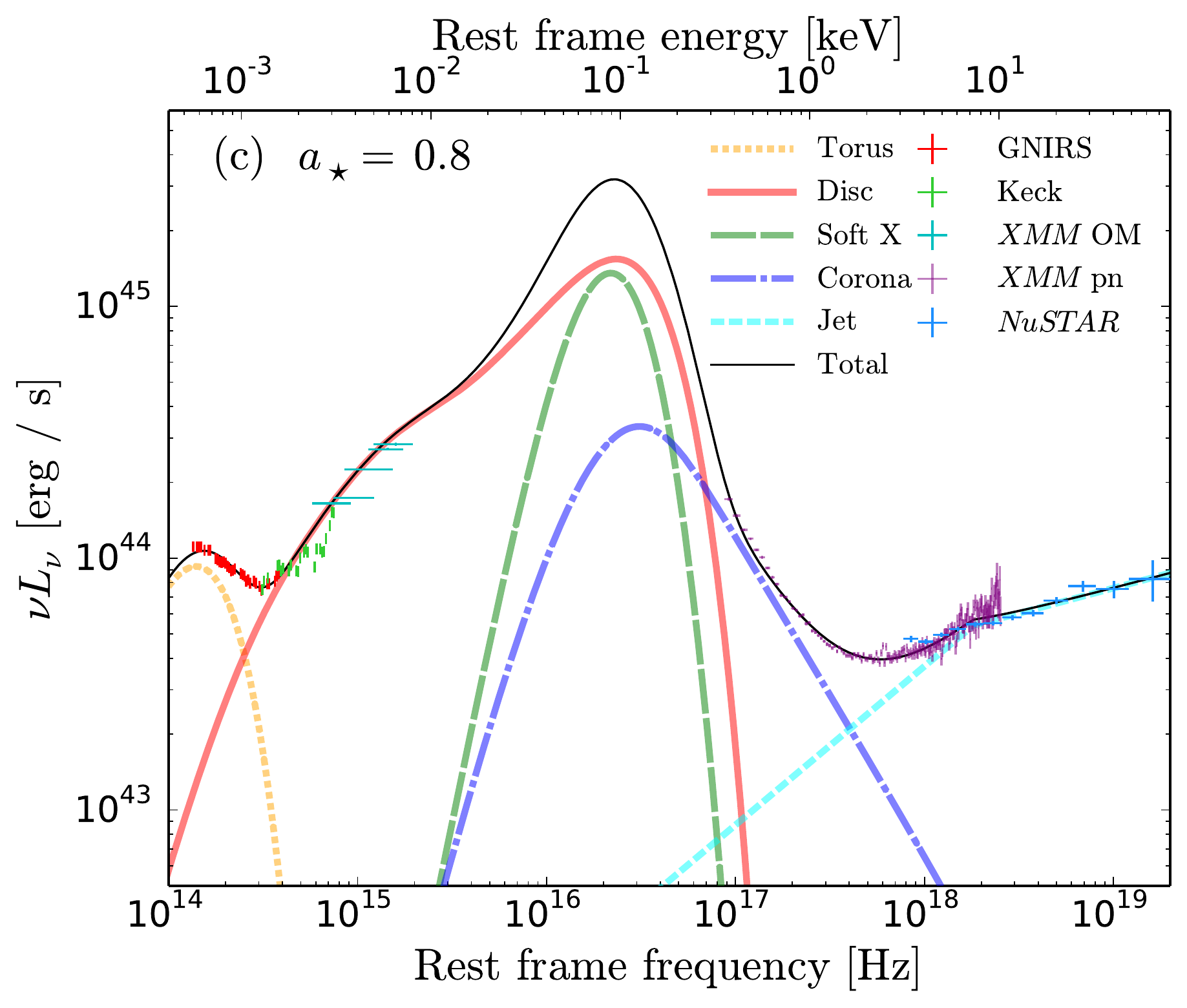} \\ 	
	\end{tabular}
\end{center}
\caption{IR to hard X-ray SEDs of 1H 0323+342. 
The data are modelled by the energy-conserving accretion model \textsc{optxconv} which calculates the emission from the accretion disc, corona and soft Comptonisation region (`Soft X').
In addition, we have added a blackbody modelling hot dust emission in the IR and, in models (b) and (c), a broken power-law to model hard X-ray emission from the jet.
See Table~\ref{tab:optxconv} in the text for the model parameters.}
\label{fig:optx}
\end{figure}

The results are presented in Table~\ref{tab:optxconv} and plotted in Fig.~\ref{fig:optx}.
Both zero spin models represent the data reasonably well, and the accretion disc and hot dust parameters are very similar.
The soft excess temperature $kT_\mathrm{e}=0.30$~keV of the zero spin, no jet model is slightly higher than is typically observed ($\left\langle kT_\mathrm{e} \right\rangle = 0.12\pm0.02$~keV, \citealt{GierlinskiDone04}).
The zero spin plus jet model shows that if the harder X-rays originate from the relativistic jet then it is possible to describe the rest of the optical-to-X-ray SED with a very typical NLS1 model.

All three models imply a relatively high Eddington ratio $L/L_\mathrm{Edd}\approx0.6$--0.8 but not super-Eddington accretion.
We estimate the accretion disc luminosity at $L_\mathrm{AD}=2.1\times10^{45}$~erg~s$^{-1}$ for the zero BH spin cases or $\approx80\%$ greater in the high-spin case.
However, the high spin model is a poorer fit to the data and cannot accommodate a soft excess component in the \textit{XMM-Newton} bandpass.
The fitting procedure lowers $kT_\mathrm{e}$ and raises $\tau$ to its maximum permitted value to force the soft excess emission out of the \textit{XMM-Newton} bandpass so as to minimise the soft X-ray power.
The soft excess thus appears to have a lower temperature than the inner accretion disc, which is unphysical in this model since the soft Comptonisation region is at smaller radii than the disc, implying that it should be hotter.
If we remove this (unseen) soft excess component by setting $f_\mathrm{pl}=1$, then we must lower $R_\mathrm{cor}$ to reduce the power in the coronal component.
Consequently, the inner radius of the accretion disc is lower and the accretion disc emission appears in the soft part of the X-ray spectrum, overpredicting the data.
In summary, an energy-conserving, high-spin model produces more soft X-ray power than is seen in the data.

With the available data we are unable to rule out the case that the corona produces all of the 2--10~keV X-ray emission. 
However, we prefer the zero spin model that includes a contribution from the jet for the following reasons.
Firstly, it gives a statistically significant improvement in the fit compared to the no jet model with $\Delta\chi^2=112$ for four additional free parameters.
Secondly, if we allow for jet emission at the hard energies, we recover parameters which are typical for a NLS1, showing a soft excess of temperature $kT_\mathrm{e}=0.22$~keV and a soft-spectrum X-ray corona.
Thirdly, the similarity of the hard X-ray photon indices (\textit{XMM-Newton}: $\Gamma_{3-10~\mathrm{keV}}=1.59\pm0.02$, \textit{NuSTAR}: $\Gamma_{3-79~\mathrm{keV}}=1.80\pm0.01$ and \textit{Swift} BAT: $\Gamma_{14-195~\mathrm{keV}}=1.73\pm0.02$) are suggestive of a single spectral component, given that the discrepancy between \textit{XMM-Newton} and \textit{NuSTAR} spectral shapes may be the result of a cross-calibration problem, as noted in \S~\ref{sec:optxconv}.
Taken together with the \textit{Fermi} data, the hard X-rays appear to be the low-energy side of the Compton hump, as we will subsequently show in our jet models.
In the following sections we proceed with the parameters determined from the model with zero BH spin plus a jet.

\begin{table*}
\small
\caption{\label{tab:opt} Results from spectral fits to the deabsorbed IR to hard X-ray SED \vspace{-2em}}
\begin{center}
\begin{tabular}{cccccccccccccccc}
\hline
\Tstrut\Bstrut
& $a_\star$ & $\nicefrac{L}{L_\mathrm{Edd}}$ & $\dot{M}$ &$R_\mathrm{cor}$ & $R_\mathrm{out}$ & $\log(L_\mathrm{AD})$ & $kT_\mathrm{e}$ & $\tau$ & $f_\mathrm{pl}$ & $\Gamma_\mathrm{cor}$ & $\log(L_\mathrm{cor})$ & $T_\mathrm{tor}$ & $\log(L_\mathrm{tor})$ & $R_\mathrm{tor}$ & $\chi^2$/dof \\ 
& & & [\nicefrac{M$_\odot$}{yr}] & [$R_\mathrm{g}$] & [$R_\mathrm{g}$] &  [erg/s] & [keV] & & & & [erg/s] & [K] & [erg/s] & [ld] &  \\
& (1) & (2) & (3) & (4) & (5) & (6) & (7) & (8) & (9) & (10) & (11) & (12) & (13) & (14) & (15) \vspace{0.1em} \\
\hline
\Tstrut
(a) & 0.0$^\mathrm{f}$ & 0.60 & 0.44 & 27.2 & 2450 & 45.27 & 0.30 & 11 				& 0.3$^\mathrm{f}$ & 1.76 & 44.53 & 1720 & 44.10 & 292 & 799/236 \\
(b) & 0.0$^\mathrm{f}$ & 0.60 & 0.44 & 24.3 & 2440 & 45.30 & 0.22 & 12 				& 0.3$^\mathrm{f}$ & 2.70 & 44.54 & 1730 & 44.10 & 297 & 687/232 \\
(c) & 0.8$^\mathrm{f}$ & 0.81 & 0.30 & 13.5 & 3380 & 45.55 & 0.03 & 100$^\dagger$	& 0.3$^\mathrm{f}$ & 3.25 & 44.83 & 1610 & 44.10 & 485 & 916/232 \\

\hline
\end{tabular}
\label{tab:optxconv}
\parbox[]{18cm}{\vspace{0.2em} The columns are: (1) dimensionless BH spin; (2) Eddington ratio; (3) mass accretion rate; (4) outer coronal radius in gravitational radii, $R_\mathrm{g}=2.95\times10^{10}$~m $=1.14\times10^{-3}$~light days; (5) outer accretion disc radius which was set to $R_\mathrm{sg}$; (6) luminosity of the accretion disc; (7) electron temperature of the soft Comptonisation region; (8) optical depth of the soft Comptonisation region; (9) fraction of the disc power below $R_\mathrm{cor}$ emitted in the power-law tail; (10) photon index of the power-law tail; (11) luminosity of the power-law tail; (12) temperature of the dusty torus; (13) luminosity of the IR radiation from the torus; (14) the dusty torus inner radius in light days, see \S~\ref{sec:tor} in the text for details; (15) the $\chi^2$ statistic over the number of degrees of freedom (dof) in the model. $^\mathrm{f}$Parameter was frozen during the fitting procedure.  $\dagger$Parameter has reached the limit of the allowed range.  $\dot{M}$ and $R_\mathrm{tor}$ are not model parameters but have been derived from our results.  These models are plotted in Fig.~\ref{fig:optx}.}
\end{center}
\end{table*}

\subsubsection{The BLR luminosity and radius}
\label{sec:BLR}
The emission from the broad line region is another important component
of the external photon field which can be Compton-scattered to higher
energies by the particles in the relativistic jet. This emission
region is located beyond the accretion disc, on a typical scale of several
light-weeks. The two relevant measures for our jet modelling are
then the luminosity and radius of the broad line region. We have
estimated the broad line region luminosity following \cite{Celotti97} as:
\begin{equation}
\label{eqn:lblr}
L_\mathrm{BLR} = \Sigma_i L_{i,\mathrm{obs}}\frac{\langle L^*_\mathrm{BLR} \rangle}{\Sigma_i L^*_{i,\mathrm{est}}},
\end{equation}  
where $\Sigma_i L_{i,\mathrm{obs}}$ is the sum of the measured
luminosities of the observed broad lines, scaled by the ratio of the
estimated total broad line region luminosity $L^*_{i,\mathrm{est}}$ to
the estimated luminosities of the observed broad lines. Both estimates
were taken from the results of \cite{Francis91} and, in the case
of H$\alpha$, from \cite{Gaskell81}. The BLR luminosity is
determined most accurately based on the actual measurement of the 
strongest emission lines, e.g.\ Ly$\alpha$, C~\textsc{iv}, H$\alpha$,
etc. Our optical spectrum covers two of the relevant broad emission lines, 
namely H$\beta$ and H$\alpha$. 
For their broad components we get a luminosity of $\log
L_{\rm H\beta} = 42.02$~erg~s$^{-1}$ and $\log L_{\rm H\alpha} =
42.44$~erg~s$^{-1}$, respectively (\citealt{Landt17}), which results
in a total BLR luminosity of $\log L_{\rm BLR} = 43.33$~erg~s$^{-1}$.

We have estimated the BLR radius in two ways, using both the near-IR
and optical radius-luminosity relationships. The near-IR
radius-luminosity relationship presented by \cite{Landt11,Landt13} is 
based on the rest-frame 1~$\mu$m continuum luminosity, which, as
these authors show, is still dominated by the ionising accretion disc
luminosity. \cite{Landt17} measured this quantity in the
near-IR spectrum to be $\log \nu L_{\rm 1\mu\mathrm{m}} =
43.92$~erg~s$^{-1}$. The derived BLR radius is then
23~light-days. The optical radius-luminosity relationship, which was
most recently calibrated by \cite{Bentz13}, using the rest-frame
5100~\AA~continuum luminosity. 
From their optical spectrum \cite{Landt17} measured this quantity
to be $\log \nu L_{\rm 5100~\AA} = 44.05$~erg~s$^{-1}$; the derived 
BLR radius is then 39 light-days. The two values resulting from the 
near-IR and optical radius-luminosity relationships are similar  
within the errors, which, when taken from the scatter in the relations,
are $\sim 40-50\%$. In the following jet modelling, we have used the 
average between the two values of 31~light-days.

We note that \cite{Wang16} calculated a BLR radius of $14.6_{-2.9}^{+7.8}$~light-days 
from the measurement of the lag in the response of H$\beta$ to changes in the continuum flux.
Whilst their estimate of the BLR radius is smaller than our two values it has a large positive 
error and is discrepant with our average value by only $\approx2\sigma$.

\subsubsection{The dusty torus luminosity and radius}
\label{sec:tor}
The dusty torus is the most extended AGN component which contributes to the
external photon field that is upscattered by the jet. 
This region is located further away from the BH than the BLR; indeed its
hottest, innermost part may be the outermost boundary of the BLR, on scales of 
light-months. The relevant measures for our jet modelling are the luminosity, 
radius and temperature of the hot dust in the torus. The luminosity and
temperature of the hot dust result directly from the blackbody fit to
the near-IR continuum and are listed in Table \ref{tab:optxconv}.
We have then estimated the hot dust radius using the theoretical relationship 
between bolometric luminosity and dust sublimation radius for grains with an 
average size given by \cite{Mor12}.
We have assumed that the dust sublimation temperature corresponds to the hot 
dust temperature and since we find this value to be $T_{\rm tor} \sim 1700$~K, 
which is much higher than the sublimation temperature of $\sim 1400$~K for 
a silicate dust composition, we have used their Eqn. (2) for pure graphite 
dust\footnote{In the case of silicate dust the radius increases by a factor of 
$\approx1.6$ compared with the value we quote here for graphite dust.}. 
The bolometric luminosity results directly from our accretion disc fits
and is listed in Table \ref{tab:optxconv}. 
The resulting hot dust radius is then $\approx300$ light-days.

\begin{table*}
\centering
	\caption{Differences between the external photon field parameters we determined and those calculated from standard assumptions}
	\label{tab:scale-epf}
	\begin{tabular}{llcccc}
	\multicolumn{2}{l}{Parameter} 				& [units] 				& Standard scaling 																				& Standard value 			& Our value \\
	\hline
 Eddington ratio		& $L/L_\mathrm{Edd}$ 		& 						& \protect{\cite{L&D11}} Eqn.~(7)$^\star$ 		 												& 0.66 						& 0.60 			 \\
 Outer corona radius	& $R_\mathrm{cor}$		& [$R_\mathrm{g}$]		&																								& 60 						& 24 			 \\
 Outer disc radius		& $R_\mathrm{out}$		& [$R_\mathrm{g}$]		&																								& 1000 						& 2440 			 \\
 Disc luminosity		& $ \log(L_\mathrm{AD})$	& [erg/s]				&																								& 45.26						& 45.30 		 \\	
 Corona luminosity		& $\log(L_\mathrm{cor})$	& [erg/s]				& $=0.1~L_\mathrm{AD}$																			& 44.26						& 44.54 		 \\	
 BLR luminosity			& $\log(L_\mathrm{BLR})$ 	& [erg/s] 				& $=0.1~L_\mathrm{AD}$  																		& 44.26  					& 43.33 		 \\
 BLR radius				& $R_\mathrm{BLR}$ 		& [$R_\mathrm{g}$] (ld) & $=5.29~R_0~\left(\frac{L_\mathrm{AD}}{10^{45}~\mathrm{erg~s}^{-1}}\right)^{\nicefrac{1}{2}}$	& $4.58\times10^4$ (52)		& $2.72\times10^4$ (30)  \\
 Dusty torus luminosity	& $\log(L_\mathrm{tor})$ 	& [erg/s]				& $=0.3~L_\mathrm{AD}$																			& 44.74						& 44.10 		 \\
 Dusty torus radius		& $R_\mathrm{tor}$ 		& [$R_\mathrm{g}$] (ld)	& $=132~R_0~\left(\frac{L_\mathrm{AD}}{10^{45}~\mathrm{erg~s}^{-1}}\right)^{\nicefrac{1}{2}}$	& $1.20\times10^6$ (1400)	& $2.63\times10^5$ (300)  \\
 Dust temperature		& $T_\mathrm{tor}$		& [K]					&																								& 370						& 1730 			 \\ 
	\hline
	\end{tabular}
	\label{tab:epf}
	\parbox[]{15.5cm}{The scaling parameter $R_0=1.89\times10^{16}$~cm.  See \cite{G&D17} for further details.
	$^\star$Here we make another measure of the Eddington ratio, scaled from the optical luminosity determined by \cite{Landt17} and assuming a radiative efficiency $\eta=6\%$ in the calculation of the bolometric luminosity $L=\eta\dot{M}c^2$.
	}
\end{table*}

\subsection{Determining the jet parameters}
\label{sec:jetmod}
\subsubsection{Calculating the observed jet emission}
\label{sec:jetcode}
\textsc{jet} is a single-zone leptonic jet emission code and based on the model presented by \citetalias{G&T09} and coded by \cite{G&D17}.
The jet is modelled as a cone with a half opening angle $\phi$ originating at the BH.    
The jet is viewed by the observer at some angle of inclination $i$. 
The model assumes that the jet emission is dominated by radiation from a single spherical `blob' of radius $R_\mathrm{diss}=\phi Z_\mathrm{diss}$ where $Z_\mathrm{diss}$ is its distance from the BH.  
The material within the jet moves with a constant bulk Lorentz factor $\Gamma_\mathrm{BLF}$.
Some fraction 
\begin{equation}
P_\mathrm{rel} = \frac{4\pi}{3}R_\mathrm{diss}^3 m_\mathrm{e} c^2 \int_{\gamma_\mathrm{min}}^{\gamma_\mathrm{max}} \gamma Q(\gamma) ~\mathrm{d}\gamma
\end{equation}
of the total jet power $P_\mathrm{j}$ is used to accelerate electrons within the emission region.
The accelerated electrons have Lorentz factors between $\gamma_\mathrm{min}$ and $\gamma_\mathrm{max}$ and injected electron population, $Q(\gamma)$, is parameterised as 
\begin{equation}
Q(\gamma) = Q_0 \left(\frac{\gamma}{\gamma_\mathrm{brk}}\right)^{-s_1} \left[1 + \left(\frac{\gamma}{\gamma_\mathrm{brk}}\right)^{s_2 - s_1}\right]^{-1},
\label{eqn:gammadist}
\end{equation}
where $s_1$ and $s_2$ are the slopes of the distribution below and above the break Lorentz factor $\gamma_\mathrm{brk}$, respectively.
These electrons then cool by both `internal' and `external' mechanisms.
The internal processes are the electrons' synchrotron emission (through interaction with the jet's magnetic field) and the Compton upscattering of these synchrotron photons by the electron population which produced them: the sychrotron self-Compton (SSC) process. 
The `external Compton' (EC) process cools electrons by the Compton upscattering of photons from the seed photon field external to the jet.  
The code runs through multiple cooling cycles until the system reaches a steady state.
The highest-energy electrons cool fastest and the steady-state distribution is found by calculating the Lorentz factor $\gamma_\mathrm{cool}$ of electrons that can just cool in the light-crossing time of the emission region and requiring this match the injected distribution below $\gamma_\mathrm{cool}$.  
    
Finally, the code calculates the observed frame emission which is boosted and blueshifted relative to the jet frame emission due to the bulk motion of the emitting plasma within the jet flow.
The observed emission from a region moving with velocity $\beta=v/c$ is a factor $\delta^3$ greater than the intrinsic emission where the Doppler factor $\delta=(\Gamma_\mathrm{BLF}[1-\beta\cos i])^{-1}$.
The code also outputs the calculated total jet power $P_\mathrm{j}=P_\mathrm{rad}+P_\mathrm{e}+P_B+P_\mathrm{p}$ which is the sum of the radiative power ($P_\mathrm{rad}$), the power in the bulk motions of electrons ($P_\mathrm{e}$) and protons ($P_\mathrm{p}$) and the Poynting power ($P_B$).  

\textsc{jet} can be used additively with the \textsc{optxconv} code by linking together the parameters $M_\mathrm{BH}$, $L/L_\mathrm{Edd}$, $i$, the comoving distance $D_\mathrm{c}$ and $z$.  
We note that this single-zone model does not calculate all of the radio jet emission.
Single-zone models calculate the emission from the base of the jet but most of the radio emission is produced further out.
Synchrotron emission in the modelled zone is strongly self-absorbed below the synchrotron self-absorption frequency, $\nu_\mathrm{ssa}$, with the spectrum below this falling off in intensity as $I\propto\nu^{5/2}$.
The observed radio slope in the SED results from the sum of emission from successive regions further along the jet with lower synchrotron self-absorption frequencies.
The \textsc{jet} code does not calculate any emission below $\nu_\mathrm{ssa}$ but in the plots shown in Fig.~\ref{fig:jet} we have subsequently added on a $\nu^{0}$ slope illustrating the emission from multiple zones and we quote $\nu_\mathrm{ssa}$ in Table~\ref{tab:jet}.

\subsubsection{The site of the $\gamma$-ray emission}
\label{sec:zdiss}
The location of the energy dissipation region $Z_\mathrm{diss}$ is an important but unknown factor in the determination of the jet SED.  
It governs the relative importance of the disc, BLR and torus seed photons in the EC process.
Disc photons always arrive from behind the jet and so are de-boosted in the jet frame.
However, because the disc is much more luminous than the BLR and torus, disc seed photons may dominate the seed photon energy density seen by the jet if the emission region is very near to the BH.
When $Z_\mathrm{diss}<R_\mathrm{BLR}$, the BLR photons are boosted in the jet frame, so the BLR component will dominate the EC seed photon energy density further from the disc where $R_\mathrm{out} < Z_\mathrm{diss} \lesssim R_\mathrm{BLR}$.
The structure and geometry of the BLR is unknown but it is modelled as a thin spherical shell.
Following \citetalias{G&T09}, the energy density of BLR seed photons is calculated in three distance ranges: interior to $R_\mathrm{BLR}$ $U^\prime_\mathrm{BLR}$ is constant (Eqn. 19 of \citetalias{G&T09}); beyond $3R_\mathrm{BLR}$ it depends on both $Z_\mathrm{diss}$ and the bulk speed of the jet (Eqn. 20 of \citetalias{G&T09}); between $R_\mathrm{BLR}$ and $3R_\mathrm{BLR}$ it is calculated as a power-law interpolation.
For a $\Gamma_\mathrm{BLF}=13$ jet, $U^\prime_\mathrm{BLR}$ decreases by more than four orders of magnitude between $R_\mathrm{BLR}$ and $3R_\mathrm{BLR}$.
When $R_\mathrm{BLR} < Z_\mathrm{diss} \lesssim R_\mathrm{tor}$ both the disc and BLR photons are de-boosted in the jet frame and the torus seed photons dominate the energy density.
  
The issue of whether $Z_\mathrm{diss}$ is near to, or far from, the BH is contentious and has been much discussed in the literature (see \citealt{MadejskiSikora16} for a recent overview).
The rapid variability of jet emission suggests a compact dissipation region. 
Under the assumptions of a conical jet that radiates across its entire cross-section, this in turn implies a dissipation region relatively near to the central engine.
On the other hand, the high energy density of UV photons near to the BH is a source of opacity to $\gamma$-rays and suggests a more distant dissipation region, particularly for objects which exhibit very high-energy (TeV) $\gamma$-ray emission.
\cite{G&T15} showed that the dissipation regions of 191 FSRQs were almost always within the BLR radius.
\cite{Sikora94} suggested that it may be Ly$\alpha$ emission from the BLR which provides the dominant source of seed photons encountered by the jet. 
Conversely, in a study of 36 FSRQ-type blazars, \cite{Zheng17} found that the dissipation regions were all outside of the BLR, and many were within the region in which the seed photon field is dominated by IR radiation from the torus.
Since we have determined the external photon field of 1H~0323+342, we can use this to predict the jet SED for a range of $Z_\mathrm{diss}$ over three orders of magnitude.
We consider the three possibilities that the seed photon field is dominated by: the accretion disc ($Z_\mathrm{diss}=1280$~$R_\mathrm{g}$: the mean $Z_\mathrm{diss}$ of FSRQs determined by \citealt{Ghisellini10}, hereafter \citetalias{Ghisellini10}); the BLR ($Z_\mathrm{diss}\approx R_\mathrm{BLR}=2.72\times10^{4}~R_\mathrm{g}$) or the torus ($Z_\mathrm{diss}\approx R_\mathrm{tor}=2.63\times10^5~R_\mathrm{g}$).
By comparing the predicted SED at each of these energy dissipation sites to the observed SED we can provide an observational constraint on $Z_\mathrm{diss}$.

\subsubsection{Constraints on input jet model parameters}
\label{sec:constraints}
Whilst $Z_\mathrm{diss}$ is \textit{a priori} unknown, we are able to fix or limit the range of several model parameters on observational or physical grounds; these are listed in Table~\ref{tab:jetfixed}.

\textit{The external photon field:}
The parameters of the external photon field are fixed to those we measured or derived from our zero spin plus jet model in \S~\ref{sec:epf}.  

\textit{The jet parameters:}
The jet viewing angle towards 1H~0323+342 was recently determined from VLBA monitoring by \cite{Fuhrmann16}.
They analysed Very Long Baseline Array (VLBA) radio images taken on several occasions between October 2010 and July 2013.
Several components in the jet had apparent velocities up to $\beta\sim7$.
Using this information the authors estimated that the jet is aligned at an angle $i\leqslant4$--$13^\circ$ to our line of sight.
If we make the reasonable assumption of $i=1/\Gamma_\mathrm{BLF}$, this also gives us a bulk Lorentz factor $\Gamma_\mathrm{BLF}\geqslant4.4$--$14.3$ which is consistent with the $\langle\Gamma_\mathrm{BLF}\rangle=13$ for FSRQs determined by \citetalias{Ghisellini10}.

To produce the observed SED slope at radio frequencies, we require the sychrotron self-absorption frequency $\nu_\mathrm{ssa}\gtrsim10^{11}$~Hz.
For $R_\mathrm{diss} = \phi Z_\mathrm{diss}$, the synchrotron self-absorption frequency
\begin{equation}
\nu_\mathrm{ssa} = \left(4.62\times10^{14}KB^{\nicefrac{5}{2}}\frac{\phi Z_\mathrm{diss}}{0.7}\right)^{\nicefrac{2}{7}}
\label{eqn:ssa}
\end{equation}
where $K$ is the normalisation of the particle distribution.
So the dominant factor governing $\nu_\mathrm{ssa}$ is the magnetic field $B$, with $\nu_\mathrm{ssa}\propto B^{\nicefrac{5}{7}}$.
The luminosity of the synchrotron peak depends on the magnetic field as $L_\mathrm{synch} \propto B^2$.
We require that the synchrotron emission does not contribute substantially to the IR part of the SED as defined by the \textit{Spitzer} and the \textit{WISE} data, which we attribute to thermal emission from the extended dusty torus.
The magnetic field must therefore be strong enough to result in a suitably high $\nu_\mathrm{ssa}$, but not so strong that the synchrotron emission dominates in the IR. 

The position and shape of the two jet emission peaks are influenced by the shape of the accelerated electron distribution.
We adopt as initial values the mean FSRQ values of Lorentz factors $\gamma$ and slopes $s$ from \citetalias{Ghisellini10}.
We leave $\gamma_\mathrm{min}$ fixed to 1 and note that the value of $\gamma_\mathrm{max}$ does not generally affect the shape of our SED substantially.

\textit{The principle of energy equipartition:}
The lowest-energy solution to jet emission requires that the electron and magnetic field energy densities are approximately equal, i.e. $U_\mathrm{e}/U_B\approx1$, (see e.g.\ \citealt{Dermer14}).
These quantities are not input parameters to the code, but they are calculated as outputs which can then be used as a check of how physically reasonable our models are.
This ratio of energy densities can be tuned if necessary by adjusting the parameters $P_\mathrm{rel}$, $B$ and $\Gamma_\mathrm{BLF}$.

\begin{table}
\small
\caption{\label{tab:jetfixed} Constraints on jet model parameters}
\begin{center}
\resizebox{\columnwidth}{!}{%
\begin{tabular}{llll}
\hline
\Tstrut\Bstrut
Param.\ 						& Value 							& Constraint 											& Ref. \\
\hline
\Tstrut\Bstrut
$M_\mathrm{BH}$ 				& $=2\times10^{7}$~M$_\odot$		& Our mass estimate										& [1] \\
$z$								& $=0.0625$							& NIR / opt.\ narrow lines								& [1] \\
\hline
$R_\mathrm{in,cor}$				& $=24.3~R_\mathrm{g}$				& Accretion disc fitting								& [\S~\ref{sec:optxconv}]  \\
$R_\mathrm{out}$				& $=2440~R_\mathrm{g}$				& Accretion disc fitting								& [\S~\ref{sec:optxconv}]  \\
$\log(L_\mathrm{cor})$ 			& $=44.54$ erg/s					& Accretion disc fitting								& [\S~\ref{sec:optxconv}]  \\
$\Gamma_\mathrm{cor}$ 			& $=2.70$							& Accretion disc fitting								& [\S~\ref{sec:optxconv}]  \\
$E^\mathrm{cut}_\mathrm{cor}$ 	& $=150$ keV						& Power-law cut-off										& [2]  \\
$\log(L_\mathrm{BLR})$ 			& $=43.33$ erg/s					& Scaled from $L_\mathrm{H\alpha,H\beta}$				& [\S~\ref{sec:BLR}]  \\
$R_\mathrm{BLR}$				& $=2.72\times10^{4}~R_\mathrm{g}$	& Scaled from $L_{\rm 1\mu\mathrm{m},5100\text{\AA}}$ 	& [\S~\ref{sec:BLR}]  \\
$\log(L_\mathrm{tor})$			& $=44.10$ erg/s					& Accretion disc fitting								& [\S~\ref{sec:tor}]  \\
$R_\mathrm{tor}$				& $=2.63\times10^{5}~R_\mathrm{g}$	& Dust sublimation radius								& [\S~\ref{sec:tor}]  \\
$T_\mathrm{tor}$				& $=1730$ K							& Accretion disc fitting								& [\S~\ref{sec:tor}]  \\
\hline
$i$								& $\leqslant4$--$13^\circ$			& Radio jet kinematics									& [3] \\
$\Gamma_\mathrm{BLF}$			& $\geqslant4.4$--$14.3$			& $i=1/\Gamma_\mathrm{BLF}$								& [3]  \\
$\phi$							& $=0.1$ radians					& Jet opening angle										& [2] \\
$B$								& $\approx2.6$ 						& $\langle$FSRQ$\rangle$ value 							& [2] \\
$\gamma_\mathrm{min}$			& $=1$								& $\langle$FSRQ$\rangle$ value							& [2] \\
$\gamma_\mathrm{brk}$			& $\approx300$						& $\langle$FSRQ$\rangle$ value							& [2] \\
$\gamma_\mathrm{max}$			& $\approx3000$						& $\langle$FSRQ$\rangle$ value							& [2] \\
$s_1$							& $\approx1$ 						& $\langle$FSRQ$\rangle$ value 							& [2] \\
$s_2$							& $\approx2.7$ 						& $\langle$FSRQ$\rangle$ value 							& [2] \\

\hline
$P_\mathrm{j}$					& $\lesssim10~L_\mathrm{AD}$		& Typical jet power										& [4]  \\
$U_\mathrm{e}/U_B$				& $\approx1$						& Equipartition											& [5]  \\
\hline
\end{tabular}}
\label{tab:jetfixed}
\parbox[]{\columnwidth}{References: [1] \cite{Landt17}; [2] \citetalias{Ghisellini10}; [3] \cite{Fuhrmann16}; [4] \cite{Ghisellini14}; [5] \cite{Dermer14}. $E^\mathrm{cut}_\mathrm{cor}$ is the high-energy cut-off of the coronal power-law; other parameters are described in \S~\ref{sec:jetmod} and \S~\ref{sec:optxconv} of the text.}
\end{center}
\end{table}

\subsection{Jet emission models}
\label{sec:jetresults}
The full SED includes low-energy data from Effeleberg / IRAM and \textit{Planck} and high-energy data from \textit{Swift} BAT and \textit{Fermi}, in addition to the mid-energy data we modelled in detail in \S~\ref{sec:epf}.
We then fit a FSRQ-like jet to our data and determine if the jet parameters we obtain are within the range found for the modelled EC emission of other blazars.
Our approach to this question is different from previous work. 
Whilst other studies of 1H~0323+342 have fit its SED including a jet (\citealt{Paliya14}; \citealt{Yao15}), they made a number of assumptions about the external photon field. 
We apply the model \textsc{bbody+optxconv+jet}, tying together the parameters $M_\mathrm{BH}$, $L/L_\mathrm{Edd}$, $i$, the comoving distance $D_\mathrm{c}$ and $z$ between \textsc{optxconv} and \textsc{jet}.
Unlike \S~\ref{sec:optxconv}, $i$ is not fixed to zero, but is set to be the inverse of the bulk Lorentz factor.

\begin{figure*}
\begin{center}
	\includegraphics[width=18cm, keepaspectratio]{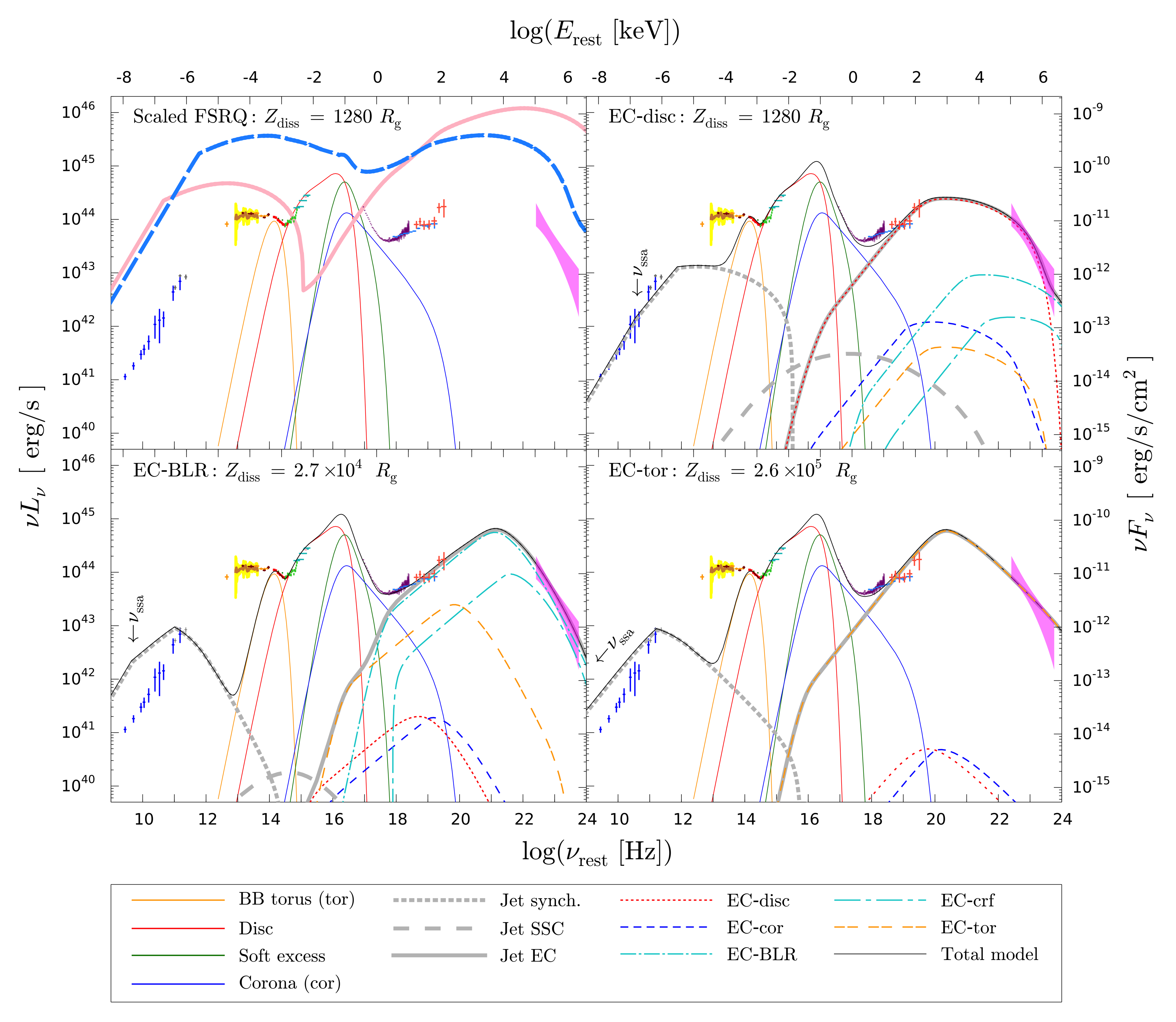}
\end{center}
\caption{SED fits to radio to $\gamma$-ray data with the jet dissipation region at increasing distance $Z_\mathrm{diss}$ from the BH. 
The model parameters are given in Table~\ref{tab:jet}.
In the top left panel the blue dashed line shows the total jet plus accretion flow emission corresponding to the scaled FSRQ model; the pink solid line shows a typical high-mass FSRQ jet scaled down in luminosity by a factor of ten.
We show the three components of the jet emission (synchrotron, synchrotron self-Compton `SSC' and external Compton `EC') as grey lines.
The individual EC components (from the disc, corona, BLR, reflection of the corona off the BLR `crf' and torus) are shown as coloured dotted and dashed lines.    
Note that the jet emission code does not calculate the radio spectrum below the synchrotron self-absorption frequency $\nu_\mathrm{ssa}$ indicated; here we have extended the radio emission to lower frequencies for illustrative purposes.
We do not model the far- or mid-IR data since we attribute this emission to the cool dusty torus; our model here includes emission from the hot dust only.
The data are colour-coded the same as in Fig.~\ref{fig:mwl}; see \S~\ref{sec:mwl} in the text for a description of the data.}
\label{fig:jet}
\end{figure*}

One might expect that the jet power $P_\mathrm{j}$ to scale with the BH mass and mass accretion rate such that $P_\mathrm{j}\propto\dot{m}M_\mathrm{BH}$ where $\dot{m}=\dot{M}/\dot{M}_\mathrm{Edd}$.
Following \cite{G&D17} we can determine $Z_\mathrm{diss}$, $P_\mathrm{rel}$ and $B$ by appropriately scaling the mean FSRQ values presented by \citetalias{Ghisellini10}.
For the other jet parameters, we adopt the mean value $\Gamma_\mathrm{BLF}=13$ and $\langle\mathrm{FSRQ}\rangle$ values for the electron distribution given in Table~\ref{tab:jetfixed}.
Applying this appropriately-scaled FSRQ jet to our external photon field gives us the scaled FSRQ model, shown as the blue line in the top-left panel of Fig.~\ref{fig:jet} and the parameters of which are given in Table~\ref{tab:jet}.
This predicted SED is very flat because $B\propto(\dot{m}/M_\mathrm{BH})^{\nicefrac{1}{2}}$ so the magnetic field for a low-mass high accretion rate object is very high (here $B=38$~G) and synchrotron cooling is highly efficient, resulting in an SED with low Compton dominance.

The product of $\dot{m}$ and $M_\mathrm{BH}$ we determine for 1H~0323+342 is a factor of ten lower than that for the average $M_\mathrm{BH}=10^9$, $\dot{m}=0.1$ FSRQ presented in \citetalias{Ghisellini10}. 
Simply scaling down the average FSRQ SED by a factor of ten produces the pink line shown in the same plot.
It is immediately apparent that whilst the product $\dot{m}M_\mathrm{BH}$ for 1H~0323+342 is an order of magnitude lower than that of a standard FSRQ, its jet luminosity is at least \textit{another} order of magnitude lower than these scalings predict.

In the EC-disc model we keep $Z_\mathrm{diss}=1280~R_\mathrm{g}$ (the same value as in the standard scaled models); as can be seen in Fig.~\ref{fig:jet}, at this location it is the disc photons which are upscattered into the Compton hump.
However, we adjust the other parameters so as to produce the best fit to the observed SED.
It is clear that it has been necessary to reduce $B$ and $P_\mathrm{rel}$ dramatically compared with the scaled FSRQ model.
As a result the total jet power $P_\mathrm{j}$ is approximately an order of magnitude lower than predicted by the scaling and in this model it is approximately half the accretion disc luminosity.
To find a near-equipartion solution it has been necessary to reduce the $\Gamma_\mathrm{BLF}$ slightly to 12, but in doing so we can achieve $U_\mathrm{e}/U_B=1.1$.
The slope $s_1$ has been increased slightly to better match the shape of the SED but the other parameters defining the accelerated electron distribution are the same.
In this model the $\gamma$-rays are produced by the upscattering of accretion disc photons, with a minor contribution from BLR photons at the hardest $\gamma$-ray energies.
The model reproduces the observed jet emission at both low and high frequencies reasonably well.

In the EC-BLR model the dissipation region has been set to $Z_\mathrm{diss}=2.7\times10^4~R_\mathrm{g}$, just inside of the BLR radius where seed photons from the BLR are responsible for the majority of the $\gamma$-ray emission.
The parameters of the accelerated electron distribution have been changed more significantly than in the EC-disc model to match the shape of the high-energy part of the SED.
However, the synchrotron component of this model now vastly overpredicts the observed radio emission.
This is partly a consequence of increasing $Z_\mathrm{diss}$ which increases the size of the dissipation region and thus reduces the energy density and lowers $\nu_\mathrm{ssa}$.
If we wish to match the high-frequency radio data in flux, we overpredict that at lower frequencies.

The dissipation region in the EC-tor model is set to $Z_\mathrm{diss}=2.6\times10^5~R_\mathrm{g}$, just inside of the hot dust radius.
The BLR emission seen by the jet is now strongly de-boosted and the distance from the BH is so great that the energy density of disc and corona seed photons is also very low. 
At this distance the seed photons from the torus are, in effect, solely responsible for the observed $\gamma$-ray and hard X-ray emission.
In the EC-tor model even with a relatively low $B$ and high $P_\mathrm{rel}$, the ratio of $U_\mathrm{e}/U_B=0.26$.
Because the jet is upscattering low-frequency photons from the torus it is necessary to increase $\gamma_\mathrm{max}$ to produce the observed $\gamma$-rays.
This EC-tor model also overpredicts the observed radio emission.

In summary, we conclude that the dissipation region must be located well within $R_\mathrm{BLR}$.

\begin{table}
\small
\caption{\label{tab:jet} Jet parameters obtained from spectral fits to the full multiwavelength SED with \textsc{bbody}+\textsc{optxconv}+\textsc{jet} models}
\begin{center}
%
\resizebox{\columnwidth}{!}{%
\begin{tabular}{lccccc}
\hline
Parameter 										& Units 				& \multicolumn{4}{c}{Model value} 						\\
\hline
\Tstrut\Bstrut												
												&						& Scaled	& EC-disc	& EC-BLR			& EC-tor			\\
$Z_\mathrm{diss}$								& [$R_\mathrm{g}$]		& 1280		& 1280		& $2.7\times10^4$	& $2.6\times10^5$	\\
$Z_\mathrm{diss}$								& [ld]					& 1.5 		& 1.5		& 30				& 300				\\
$a_\star$										&						& 0.0		& 0.0		& 0.0				& 0.0	  			\\
$i$												& [deg]					& 4.41 		& 4.77		& 4.98				& 4.98  			\\
$\Gamma_\mathrm{BLF}$							&						& 13.0		& 12.0		& 11.5				& 11.5  			\\
$\delta$										&						& 13.0 		& 12.0		& 11.5				& 11.5 				\\
$B$												& [G]					& 38.0 		& 8.00		& 0.75				& 0.15  			\\
$\gamma_\mathrm{min}$							&						& 1.00 		& 1.00		& 1.00				& 1.00 	 			\\
$\gamma_\mathrm{brk}$							&						& 300 		& 300		& 150				& 300  				\\
$\gamma_\mathrm{max}$							& 						& 3000 		& 3000		& 3000				& 30000			   	\\
$\gamma_\mathrm{cool}$							&						& 19 		& 47		& 58				& 163   			\\
$s_1$											&						& 1.00 		& 1.50		& 2.00				& 1.50 				\\
$s_2$											&						& 2.70 		& 2.70		& 4.25				& 3.20  			\\
$\log(\nu_\mathrm{ssa})$						& [Hz]					& 11.6 		& 10.6		& 9.67				& 8.76				\\
$\log\left(\nu_\mathrm{peak}^\mathrm{sync}\right)$			& [Hz]		& 13.8 		& 12.5		& 11.0				& 11.2  			\\
$\log\left(\nu L_{\nu_\mathrm{peak}}^\mathrm{sync}\right)$	& [erg/s]	& 45.56 	& 43.12		& 42.96				& 42.92  			\\
$\log(P_\mathrm{rel})$							& [erg/s]				& 42.24 	& 41.00		& 41.80				& 41.50				\\
\hline
$\log(P_\mathrm{rad})$							& [erg/s]				& 43.95 	& 42.51 	& 42.76 			& 42.74  			\\
$\log(P_\mathrm{e})$							& [erg/s]				& 43.76 	& 42.74		& 43.53				& 43.23  			\\
$\log(P_B)$										& [erg/s]				& 44.12 	& 42.70		& 43.25				& 43.82  			\\
$\log(P_\mathrm{p})$							& [erg/s]				& 45.91 	& 45.01		& 46.14				& 45.32  			\\
$\log(P_\mathrm{j})$							& [erg/s]				& 45.93 	& 45.01		& 46.14				& 45.34  			\\
$P_\mathrm{j}/L_\mathrm{AD}$					& 						& 4.3 		& 0.52		& 6.9				& 1.1	  			\\
$U_\mathrm{e}/U_B$								&						& 0.44 		& 1.1		& 1.9				& 0.26  			\\

\hline
\end{tabular}
}
\label{tab:jet}
\parbox[]{9.75cm}{\vspace{0.2em} Parameters are described in \S~\ref{sec:jetcode} and \S~\ref{sec:jetresults} of the text.}
\end{center}
\end{table}

\section{Discussion}
\label{sec:disc}
\subsection{Is 1H~0323+342 a typical NLS1?}
The 2--10~keV photon indices of NLS1s are generally soft ($\langle\Gamma\rangle_\mathrm{NLS1}=2.19\pm0.10$, \citealt{Leighly99}), whereas that of 1H~0323+342 is much harder ($\Gamma_{2-10~\mathrm{keV}}=1.7$).
The X-ray RMS spectrum of the fast variability shows a clear break at $\approx1$~keV.
At least two spectral components are therefore required to fit the \textit{XMM-Newton} EPIC spectra.
\cite{Landt17} did not find such clear evidence for multiple spectral components in their analysis of \textit{Swift} XRT data.
Curvature in the X-ray spectrum was only apparent in the co-added spectrum of three \textit{Swift} observations.
Here, the higher quality of X-ray data obtained from a long \textit{XMM-Newton} observation affords us a better opportunity for a more detailed spectral decomposition.
However, some degeneracy between spectral models still remains, as was discussed in \S~\ref{sec:optxconv} where we presented energy-conserving, physical models of the X-ray spectra.

Previous studies have attempted to determine the BH spin of 1H~0323+342 by fitting a blurred reflection model to X-ray spectral data, but the results are not conclusive.
\cite{Paliya14} found a high spin with $a_\star=0.96\pm0.14$ by modelling \textit{Swift} XRT and BAT data whereas \cite{Yao15} found an upper limit of $a_\star<0.13$ using \textit{Suzaku} data of a more limited energy range.
We explored models which included the effects of BH spin in \S~\ref{sec:optxconv}.
Both zero spin models provide reasonably good fits to the data, but the high spin model overpredicted the soft X-ray power.
Our modelling is therefore suggestive of a low BH spin scenario for 1H~0323+342.
However, our model makes a number of assumptions which, if relaxed, could in principle allow for a higher BH spin.
The model is energy-conserving and assumes that the accretion power passing through the outer and inner disc is equal.
This would not be the case if some power were lost as e.g.\ as disc wind or transported up the jet itself (\citealt{BP82}).
A recent well-studied example where this may be the case is the super-Eddington AGN RX~J0439.6-5311 (\citealt{Jin17a,Jin17b}).
A larger BH mass could also allow for a higher spin.
Here, we fixed the mass to the value $M_\mathrm{BH}=2\times10^7$~M$_\odot$ determined by \cite{Landt17} from measurements of the hydrogen Balmer and Paschen lines. 
However, there is emerging evidence (from e.g.\ accretion disc peak fitting, \citealt{Calderone13}; spectropolarimetry, \citealt{Baldi16} and the $M_\mathrm{BH}$-$L_\mathrm{bulge}$ relation, \citealt{DAmmando17}) that the BH masses of NLS1s are underestimated when using the standard single-epoch virial methods.
These studies have found NLS1 BH masses more in line with the rest of the RL-AGN population with $M_\mathrm{BH}\geqslant10^8$~M$_\odot$.
In the case of this source, the two-month reverberation mapping study of 1H~0323+342 by \cite{Wang16} also found a similarly low BH mass of $\left(3.4_{-0.6}^{+0.9}\right)\times10^7$~M$_\odot$, so at present there is no strong evidence that the BH mass of 1H~0323+342 is substantially greater than the value we have used.
Our model also assumes that the accretion flow is not disrupted by the launching and presence of the powerful relativistic jet.

When modelling the \textit{XMM-Newton} X-ray spectra, we allowed the Galactic column to be a free parameter and found the best fits required an excess of Galactic column density of about 50\% above the \citetalias{DL90} value.
Whilst the modelled value is more similar to the total column quoted by \cite{Willingale13}, we consider that it is unlikely that the Galactic column is in fact as high as our models determine since we found no evidence for the additional absorption in the \textit{XMM-Newton} RGS spectrum or our optical and UV data (see the Appendix for further details). 
However, it is well established that NLS1s commonly exhibit complex intrinsic absorption (e.g.\ \citealt{Komossa00}) and emission (e.g.\ \citealt{Smith&Page08}) features.
We tested several possibilities including neutral or ionised intrinsic absorption, and \textit{ad hoc} absorption and emission features following \cite{Gallo04} but these did not make as great an improvement in the fit as the increased neutral Galactic column or `correct' the shape of the deabsorbed spectrum.
We have adopted the increased Galactic column as the simplest solution in our models which corrects the spectral shape and gives the greatest improvement in the fit statistic.
Doing so does not substantially change the main conclusions from our subsequent SED modelling.
However, if we do not allow for this additional absorbing column, then we are unable to include a soft Comptonisation region in the models presented in \S~\ref{sec:optxconv}.  
Consequently the corona photon index is slightly harder and the accretion disc emission increases by a factor $\approx1.4$ compared with our zero spin plus jet model.
Nevertheless, these changes are not significant enough to basically alter our jet modelling or conclusions.

Apart from its jet, there is nothing we have found here that sets 1H~0323+342 apart from other NLS1s; its mass and Eddington ratio are both only slightly higher than the average values reported by \cite{Rakshit17}.
Our extensive exploration of the modelling parameter space shows that the IR-to-X-ray SED of 1H~0323+342 is within the range observed for NLS1s, but with the addition of a jet component.
Why this particular NLS1 possesses a relativistic jet when the vast majority of others do not therefore remains an open question.

\subsection{Contribution of the jet to the IR and X-ray emission}
\label{sec:irx}
Turning to a much lower-frequency part of the SED, our \textit{Spitzer} IRS spectrum is shown in Fig.~\ref{fig:Spitzer}.
It can been seen in the figure that the \textit{WISE} photometry agrees well with our \textit{Spitzer} data despite the former being taken approximately eighteen months later, indicating very little, if any, variability over a timescale of years.  
In the \textit{Spitzer} IRS spectrum we can see two strong, broad humps at $\approx11$ and 18~$\mu$m which we attribute to the 9.7 and 18~$\mu$m silicate features commonly seen in emission in type 1 AGN spectra.
The 9.7~$\mu$m feature is often not observed at the rest-frame wavelength 9.7~$\mu$m, but redward of this position.
The apparent redshift of this feature was seen in all of the \textit{Spitzer} spectra of a sample of 12 RL-AGN studied by \cite{Landt10}, and had previously been seen in a few other sources (e.g.\ \citealt{Schweitzer08}; \citealt{Sturm05}; \citealt{Siebenmorgen05}) but its cause is currently unknown.
Some weak polycyclic aromatic hydrocarbon (PAH) features may also be present in the spectrum.
Since we do not see a featureless continuum or flux variability which we would expect from synchrotron emission,
these features suggest that most of this mid/far-IR emission originates from the torus rather than from the jet.
This interpretation is contrary to that of others (\citealt{Abdo09b}; \citealt{Paliya14}; \citealt{Yao15}) who have studied this object and attributed the IR emission to the jet synchrotron component.
As a consequence our jet models have a lower-luminosity synchrotron peak and higher Compton dominance in the SED.

In \S~\ref{sec:spitzer} we measured the flux and hence the luminosity of the mid-IR emission line [O~\textsc{iv}]~$\lambda25.89~\mu$m.
We compared the luminosity of the line to the X-ray luminosities in the \textit{XMM-Newton} and \textit{Swift} BAT bands.
Using the derived luminosity of [O~\textsc{iv}], $\log(L_\mathrm{[O\textsc{iv}]})=41.3$, and the 2--10~keV X-ray luminosity, 
$\log(L_{2-10~\mathrm{keV}})=43.9$, we can compare these values with those of the sample of AGN studied by \cite{Diamond-Stanic09}; we see that 1H~0323+342 is broadly consistent with other Seyfert 1s shown in their Fig.~4.
However, if we now look at the corresponding relation between the hard X-rays and [O~\textsc{iv}] (e.g.\ \citealt{LaMassa10}, their Fig.~8) we see that 1H~0323+342, having $\log(L_{14-195~\mathrm{keV}})=44.45\pm0.05$, lies above the correlation, suggesting it is more luminous in hard X-rays by a factor of about five with respect to the other Seyfert 1 AGN in their sample.
The detection of the [O~\textsc{iv}] line is clearly of limited significance, given the S/N of the \textit{Spitzer} IRS spectrum.
If instead we treat the measured flux as an upper limit, then 1H~0323+342 must be even more overluminous in 14--195~keV X-rays, in terms of the correlation found for other AGN.
This excess of hard X-ray luminosity supports our preference for a model in which emission from the relativistic jet makes a contribution to the hard X-rays.

In \S~\ref{sec:FeK} we show that the continuum fit to the X-ray spectrum is improved if we add a narrow line feature at 6.43~keV with $\mathrm{EW}=36\pm8$~eV which we associate with neutral Fe~K$\alpha$.
For comparison, \cite{Shu10} measured the EWs of the narrow cores of Fe~K$\alpha$ in a sample of Seyfert AGN observed by \textit{Chandra}, finding $\langle\mathrm{EW}\rangle=53\pm3$~eV.
In RL AGN, the contribution to the X-ray continuum of Doppler-boosted emission from the jet will result in a relative weakness (of the EW, by dilution from the additional continuum flux) of the fluorescent Fe~K$\alpha$ emission line. 
\cite{Bianchi07} derived a relationship between the EW of the narrow Fe~K$\alpha$ and the 2--10~keV luminosity based on RQ type 1 AGN (Eqn.~(1) in their paper).
From this relation, we can estimate that the narrow Fe~K$\alpha$ EW should be $\approx56$~eV if 1H~0323+342 were RQ and the jet made no contribution to the 2--10~keV continuum.
The lower EW we have determined tentatively suggests some jet emission may be present in the \textit{XMM-Newton} bandpass.

\subsection{The external photon field}
Our new approach here was to use the wealth of quasi-simultaneous spectroscopy and photometry to derive the seed photons for the external Compton components input into the jet code. 
This is clearly a better approach than assuming a given SED shape, especially given our well-sampled SED.

The accretion disc luminosity (consequently the Eddington ratio) we determine from our zero spin plus jet model actually agrees very well with the values we would obtain from estimating the mass accretion rate $\dot{M}$ from the optical luminosity (\citealt{L&D11}) and assuming $\eta=6\%$ to calculate the bolometric luminosity from $L=\eta\dot{M}c^2$.
The agreement of these values with our zero spin plus jet SED model lends some support to this model over the high-spin case model where the Eddington ratio and disc luminosity were both higher.
In the \citetalias{G&T09} model, the corona extends out to 60~$R_\mathrm{g}$ and has one tenth of the accretion disc luminosity.
As can be seen from Table~\ref{tab:epf}, our corona is more luminous but more compact than this, extending to 24~$R_\mathrm{g}$.
The corona photon index in the \citetalias{G&T09} model is assumed to be 2 whereas the values we determined are much softer at 3.59 in \S~\ref{sec:xspec} and 2.7 in the zero spin plus jet model.
Despite being more luminous than the standard model assumes, the EC-corona emission is not a very strong component in any of our models.
The \textsc{jet} model is insensitive to subtle changes in the spectral shape and geometry of the corona so the difference in photon indices and radii also have very little impact on our results.
\citetalias{G&T09} scale the BLR luminosity and radius from the accretion disc luminosity.
Table~\ref{tab:epf} shows that their standard assumptions predict a larger and much more luminous BLR than we determined in \S~\ref{sec:BLR}.    
The smaller radius we determine is in better agreement with the value obtained from the reverberation mapping study of \cite{Wang16}.   
Interestingly, our value for the BLR luminosity is a factor of ten lower than the \citetalias{G&T09} model assumption that it is one tenth of the accretion disc luminosity. 
The BLR radius that we determine is just over half of the value calculated in the standard model.
However, the energy density of BLR seed photons is $U'_\mathrm{BLR}\propto L_\mathrm{AD}$ and $U'_\mathrm{BLR}\propto R_\mathrm{BLR}^{-2}$, so the overall difference in $U'_\mathrm{BLR}$ is a factor $\approx3$ which does not substantially change our conclusions.

The torus we adopt is smaller and less bright than the standard model assumes.
Our infrared spectrum only samples emission from the hottest dust on the inner edge of the torus, so the temperature we determine from our models is much greater than that in the \citetalias{G&T09} prescription which characterises the dust as much cooler and more extended.
Of course, both are simplifications of the actual torus temperature-radius and luminosity-radius profiles.
Had we used the standard assumptions with an energy density $U'_\mathrm{tor}$ smaller by a factor $\approx5$, the torus component in the EC-BLR model would be weaker.
The dissipation region in our EC-tor model would have been placed even further out, so $v_\mathrm{ssa}$ would be smaller and the model would still overpredict the radio emission. 

\subsection{The impact of variability}
\label{sec:dvar}
We have reason to claim that the non-simultaneity of our broadband data does not strongly affect our results and conclusions.
The \textit{XMM-Newton} optical/UV photometry and X-ray spectra are truly simultaneous and sample the outer and inner accretion flows, respectively.
The high-frequency end of the \textit{XMM-Newton} spectra and low-frequency end of the \textit{NuSTAR} spectra are of very similar flux levels despite the seventeen-month gap between observations.  
As noted in \S~\ref{sec:optxconv}, the apparent discrepancy in spectral shape may be due to a calibration issue.
Considering the higher frequencies, the flux levels of the \textit{NuSTAR} and \textit{Swift} BAT X-ray spectra are consistent in their region of overlap and the \textit{Fermi} $\gamma$-ray data was chosen to sample the period covering the \textit{XMM-Newton} observation.  
Turning to lower frequencies, the \textit{XMM-Newton} optical photometry (simultaneous with the X-rays) are consistent with points sampling the continuum determined from the Keck optical spectrum.
The GNIRS spectrum was flux-scaled to match the Keck data, because we suspect that the apparent difference in flux is due to a shift resulting from the uncertain absolute flux calibration of the near-IR spectrum rather than genuine source variability (as described in \citealt{Landt17}).
After this correction has been applied, the near-IR data then appears to connect with the mid- and far-IR bands sampled by \textit{Spitzer} and \textit{WISE}.  As we commented in \S~\ref{sec:irx}, the \textit{WISE} photometry agrees very well with the \textit{Spitzer} data even though the observations were separated by more than a year.

With regards to the Effelsberg and IRAM radio data, \cite{Angelakis15} reported flux density variability magnitudes on average $\approx30\%$ and up to 63\%, with the variability being more pronounced at higher frequencies.
Whilst the flux densities at lower frequencies were generally stable, frequencies 14.6~GHz and above exhibited occasional flaring.
However, the mean values from which we used in our analysis are not strongly affected by the flaring episodes and are broadly consistent with the stable, baseline level.

We noted in \S~\ref{sec:med-var} and \S~\ref{sec:Fermi} that the X-ray and $\gamma$-ray data were obtained during periods of low activity.
Given the constancy in flux betweeen neighbouring frequency bands, it is therefore reasonable to conclude that all of our multiwavelength data is appropriate to describe this source in a low state.
Although our data are mostly not simultaneous, for the reasons given above we do not expect that this impairs our overall conclusions.

\subsection{The origin of the $\gamma$-ray emission}
It is generally accepted that in the case of a high-accretion rate blazar, such as 1H~0323+342, the $\gamma$-ray emission from 1H~0323+342 results from the EC process.
Here we compare our general findings with those from other similar studies.
Both \cite{Abdo09b} and \cite{Paliya14} (for the quiescent state) found that the dissipation region must be relatively near to the BH, with $Z_\mathrm{diss}\approx1300~R_\mathrm{g}$, very similar to the value we used in the EC-disc model here, which adopts the mean FSRQ $Z_\mathrm{diss}$ of \citetalias{Ghisellini10}.
\cite{Yao15} were unable to constrain the location so well, since models with a dissipation region located inside or outside of the BLR both reproduced their broadband SED reasonably well.
In our preferred EC-disc model, the jet emission region appears to be relatively near the accretion disc, with EC-disc photons producing the hard X-rays and the $\gamma$-rays.
This is different to the findings of \cite{Paliya14} where EC-BLR photons were dominant in all states (both quiescent and flaring).
This is also different to \cite{Yao15} who considered only EC-BLR and EC-torus situations.
Here, both the EC-BLR and EC-tor models are shown to overpredict the observed radio emission.

In many jet models, such as ours, it is assumed that $R_\mathrm{diss}=\phi Z_\mathrm{diss}$, therefore a compact emission region (with small $R_\mathrm{diss}$) must be relatively near to the core of the AGN.
However, other geometries have been proposed such as the `spine-sheath' (e.g.\ \citealt{Sol89}, \citealt{GTC05}, \citealt{Sikora16}) or `turbulent cell' (e.g.\ \citealt{Marscher&Jorstad10}) models.
In these cases, the jet does not radiate across its entire cross-section so a compact emission region does not necessarily imply one that is close to the BH.
For simplicity, and for the ease of comparison with the work of other authors, we have not considered alternative jet geometries here.

Overall, our EC-disc model has a set of parameters that best match the broadband SED, across an exceptionally wide range of frequencies, from the radio to $\gamma$-rays. 
This model has the added attraction that is very close to an energy equipartition solution with $U_\mathrm{e}/U_B=1.1$. 
The parameters of the accelerated electron distribution for this model are the same as those for the scaled-down FSRQ model, with the exception of the slope $s_1$ which is 1.5 in the former and 1.0 in the latter. 
It differs from the scaled FSRQ model mainly in that its magnetic field and power injected into the electrons are much lower than predicted; we discuss this further in the next section.    

\subsection{Where does 1H~0323+342 lie in the blazar sequence?}
Our interpretation of mid-IR emission as being torus-dominated limits the peak luminosity of the synchrotron component and increases the dominance of the EC peak in the SED.
We therefore arrive at an SED shape typical of a FSRQ but at a luminosity more like that of a BL Lac.
It has been predicted that low-mass, lower-luminosity FSRQs would be detected by \textit{Fermi}, which has a greater sensitivity than its predecessor \textit{EGRET}.
However, the blue line in the top-left panel of Fig.~\ref{fig:jet} shows that a scaled-down FSRQ SED is both more luminous and has a flatter shape more like a BL Lac than is observed.
If we simply scale down a typical FSRQ SED by a factor of ten (the pink line in the top-left panel of Fig.~\ref{fig:jet}), it is also much more luminous than the data although the shape is more similar to the one we fit in the EC-disc model.  
In both cases the synchrotron and Compton humps in the SED are at frequencies more typical of FSRQs than the `bluer' SEDs of BL Lacs.  The accelerated electron distribution in our EC-disc model has parameters very similar to that of a typical FSRQ; the higher-frequency peaked BL Lacs have much greater $\gamma_\mathrm{brk}$ and $\gamma_\mathrm{max}$.
Additionally, the bulk Lorentz factor of this model is more similar to that of an FSRQ than a BL Lac (which have $\langle\Gamma_\mathrm{BLF}\rangle=15$, \citetalias{Ghisellini10}). 

\cite{Ghisellini14} found a clear positive correlation between the jet and disc powers in a sample of over 200 blazars.
As well as this relation, they also found that the jet powers exceeded the accretion disc luminosities typically by a factor $\sim10$.
Clearly, this is not the case for 1H~0323+342 where the jet power in our EC-disc model is approximately half the disc luminosity.
Zero BH spin implies a low radiative efficiency, $\eta=0.06$, and we can determine that $\log\left(\dot{M}c^2\right)=46.5$, so 1H~0323+342 lies well outside of the $3\sigma$ dispersion of $P_\mathrm{j}$-$\dot{M}c^2$ determined by \cite{Ghisellini14}.
Even if we allow for a high spin (which our energy-conserving models disfavoured) we calculate $\log\left(\dot{M}c^2\right)=45.8$ and 1H~0323+342 is then only just inside of the 3$\sigma$ region.
We showed in our EC-disc jet model that in order to match the observed SED it is necessary to reduce both $P_\mathrm{rel}$ and $B$ from the values predicted by the scaled FSRQ model.
As well as having a very low jet power for a FSRQ, 1H~0323+342 has a low jet power compared with the prototypical $\gamma$-NLS1 PMN~J0948+0022, as was noted by both \cite{Abdo09b} and \cite{Paliya14}.
Since the strength of the magnetic field determines how efficiently the jet can extract the rotational energy of the BH, it is possible that the (relatively) weak magnetic field of 1H~0323+342 is less well able to extract spin power and inject it into the jet.  
Our findings also indicate that it is plausible that 1H~0323+342 has a lower BH spin than other blazars, and consequently is unable to host as powerful a jet.

\subsection{A comparison of jet powers}

\begin{table}
\small
\caption{Comparison of 1H~0323+342 jet powers}
\begin{center}
\resizebox{\columnwidth}{!}{%
\begin{tabular}{lccccc}
\hline
Jet model 				& $\log(P_\mathrm{rad})$	& $\log(P_\mathrm{e})$ 	& $\log(P_B)$	& $\log(P_\mathrm{p})$ 	& $\log(P_\mathrm{j})$ 	\\
						& (1)				& (2)				& (3)	& (4)				& (5)				\\
\hline
EC-disc ($5^\circ$)		& 42.51				& 42.74 			& 42.70 & 45.01				& 45.01 			\\
EC-disc ($3^\circ$)		& 42.18				& 42.15 			& 42.58 & 44.44				& 44.45 			\\
Abdo$^\mathrm{a}$		& 42.8				& 42.7  			& 43.3  & 44.3				& 44.4  	\\
Paliya$^\mathrm{b}$		& 41.29				& 					& 		& 44.06				&       	\\
Yao$^\mathrm{c}$		& 43.9				& 43.4 				& 42.6 	& 43.7				& 44.2 		\\
\hline		 
\end{tabular}
} 
\label{tab:pow}
\parbox[]{\columnwidth}{\vspace{0.2em} Here we compare the jet powers calculated for our EC-disc model with those of: $^\mathrm{a}$\cite{Abdo09b}; the quiescent state model of $^\mathrm{b}$\cite{Paliya14} and the IC/BLR model of $^\mathrm{c}$\cite{Yao15}.
In the columns we quote the logarithms of: (1) the radiative power; (2) the power in the bulk motion of electrons; (3) the Poynting power; (4) the power in the bulk motion of protons and (5) the total jet power, in units erg~s$^{-1}$.}
\end{center}
\end{table}

We claim that 1H~0323+342 hosts an underpowered jet for a FSRQ, compared with those presented by \citetalias{Ghisellini10} and \cite{Ghisellini14}.
The jet power that is determined is strongly dependent on the assumptions made in the modelling.
Other authors have determined the jet power of 1H~0323+342 by fitting a single-zone leptonic jet model to its broadband SED; we tabulate the relevant values in Table~\ref{tab:pow}.
It can be seen that power that was calculated for our preferred model, `EC-disc ($5^\circ$)', is greater than those of these previous studies and here we discuss some of the differences.

The most straightforward comparison is to the model adopted by \cite{Abdo09b} because they use the most similar modelling prescription to our own.
However they have adopted a BH mass estimate half of our value, so whilst their $R_\mathrm{diss}$ is equal to ours in mass-scaled units, it is a factor of two smaller in absolute terms which affects the calculated energy densities. 
Another key difference is their use of a smaller inclination angle $i=3^\circ$ rather than our value of $i=1/\Gamma_\mathrm{BLF}\approx5^\circ$, although they use the same $\Gamma_\mathrm{BLF}=12$ as us.  
The Doppler boosting in their case is therefore greater by a factor of four and they can fit the observed $\gamma$-ray emission with a jet which is around five times less powerful than ours.
We find that we can replicate the shape of our EC-disc SED model at a lower inclination angle of $3^\circ$ by turning down $B$ and $P_\mathrm{rel}$, but keeping $\Gamma_\mathrm{BLF}=12$.  
In this case we obtain a jet power very similar to \cite{Abdo09b}, as shown in Table~\ref{tab:pow}.

The quiescent state model of \cite{Paliya14} has approximately an order of magnitude lower kinetic power than our model.
This difference is in part due to their choice of a much lower $\Gamma_\mathrm{BLF}=7$; 
since $P_\mathrm{p}\propto\Gamma_\mathrm{BLF}^2$, for the same number of protons the kinetic power would be reduced by a factor $\approx0.3$.

The IC/BLR model of \cite{Yao15} has a very low $\Gamma_\mathrm{BLF}=2.7$, therefore the bulk motion of particles is not the dominant factor in the jet power, and the radiative power contributes approximately half of the total jet power. 
Their injected electron distribution is skewed towards higher Lorentz factors, with $\gamma_\mathrm{brk}=1073$ in their case compared with our value of $\gamma_\mathrm{brk}=300$.
As a result, the power in the bulk motion in protons is only approximately twice the power in the bulk motion in electrons.  
However, since they do not quote the injected power $P_\mathrm{rel}$ we are unable to make a more detailed comparison.

This diversity of jet powers illustrates the strong dependence on the modelling assumptions.
Since we adopted the same approach as \citetalias{Ghisellini10}, the most appropriate comparison is to their large sample of FSRQs.
The models of other authors can fit similar SEDs for this source and they have found even lower jet powers.
Therefore, we are confident that our principal conclusion that 1H~0323+342 hosts a low-powered jet remains robust. 

\section{Summary and conclusions}
\label{sec:conc}
We assembled a well-sampled and wide-ranging multiwavelength data set including our new infrared, optical and X-ray spectra and supplemented these with archival data including spectra and photometry in other wavebands from radio through to $\gamma$-rays.
The observations, data reduction and reference sources were described in \S~\ref{sec:mwl}.
In \S~\ref{sec:xray} we performed a temporal and spectral analysis of a long (80~ks) \textit{XMM-Newton} observation.
We found evidence for complexity in the low-energy range of the X-ray spectrum which is possibly due to absorption in addition to the Galactic column. 
The dereddened / deabsorbed IR, optical and X-ray spectra and optical/UV photometry were used to fit an energy-conserving accretion disc model to our data in \S~\ref{sec:optxconv}.
The results from this modelling, along with measurements of emission lines observed in our optical spectrum, allowed us to define the photon field in the vicinity of the central engine of 1H~0323+342.  
In particular, we determined the size scales and luminosities of the accretion disc and its corona, the BLR and the dusty torus.
We then introduced these parameters into a relativistic jet emission code to determine the jet parameters which best reproduce the observed SED.
The results from our modelling of the jet are presented in \S~\ref{sec:jetresults}.

Our main conclusions are as follows:
\begin{enumerate}[(i)]
\item It is possible to fit an energy-conserving accretion flow model to the IR-to-X-ray SED in which the accretion flow has parameters typical of a NLS1 and where the jet makes a contribution to the hard X-rays.
This is only possible if the BH spin is low or zero; a high BH spin model predicts more energy in soft X-rays than is seen in the data.
We find the X-ray emission has contributions from a soft-spectrum corona and a soft Comptonisation region with temperature $kT_\mathrm{e}=0.22$~keV, and determine a relatively high Eddington ratio of $L/L_\mathrm{Edd}=0.6$.  
\item We detect a weak iron line in the \textit{XMM-Newton} EPIC spectra which has an energy consistent with neutral Fe~K$\alpha$ fluorescence.
\item We find that 1H~0323+342 has a broadband SED with a similar shape to an FSRQ (showing high Compton dominance) but with a similar luminosity to a BL Lac.
We show that this source is \textit{not} a consistent with being a mini FSRQ, since scaling down standard FSRQ jet parameters by BH mass and mass accretion rate produces an SED model which vastly overpredicts the observed emission.
The jet in 1H~0323+342 appears to be underpowered by at least an order of magnitude compared with predictions made by scaling an average FSRQ jet.
With respect to the accretion power, the source lies outside of the 3$\sigma$ dispersion region of the $P_\mathrm{j}$-$\dot{M}c^2$ relation determined by \cite{Ghisellini14}.
\item We show that (within the assumptions of our jet model) the energy dissipation region of the jet must be located near to the BH and well within the BLR radius.
In our preferred jet emission model, seed photons from the accretion disc are upscattered to produce the observed $\gamma$-ray emission.
\end{enumerate}

Our detailed study of 1H~0323+342 has shed new light on its accretion properties e.g.\ the Eddington ratio the nature of its outflow (jet), the interplay between the relativistic particles and the radiation field and its relation to other blazars.  
However, this is only one example of the small group of $\gamma$-NLS1s, and in-depth studies of a number of other examples need to be made to reveal whether they share similar characteristics, or are a heterogeneous sample. 

\section*{Acknowledgements}
DK acknowledges the receipt of an STFC studentship (ST/N50404X/1). 
DK, HL, MJW, CD and EG acknowledge support from the STFC (ST/L00075X/1).
MPS acknowledges support from STFC through grant ST/N000919/1, the John Fell Oxford University Press (OUP) Research Fund and the University of Oxford.
Thanks goes to both Emanuele Nardini and Filippo D'Ammando for useful discussions regarding the X-ray spectral analysis.

In this paper we have made use of the following:
\begin{itemize}
\item data from \textit{Fermi}, a NASA mission operated and funded NASA, the U.S. Department of Energy and institutions in France, Germany, Japan, Italy and Sweden;
\item data from \textit{Spitzer}, \textit{WISE}, \textit{Planck} and 2MASS which are available from the NASA / IPAC Infrared Science Archive, which is operated by the Jet Propulsion Laboratory, California Institute of Technology, under contract with NASA;
\item data from \textit{NuSTAR} a project led by the California Institute of Technology, managed by the Jet Propulsion Laboratory and funded by NASA;
\item data from \textit{Swift}, and its XRT Data Analysis Software (XRTDAS) developed under the responsibility of the ASI Science Data Center (ASDC), Italy;
\item data from and software developed for \textit{XMM-Newton}, an ESA science mission with instruments and contributions directly funded by ESA Member States and NASA; 
\item data and software (including the ftools\footnote{\url{http://heasarc.gsfc.nasa.gov/ftools/}} \citealt{Blackburn95}) provided by the High Energy Astrophysics Science Archive Research Center (HEASARC), which is a service of the Astrophysics Science Division at NASA/GSFC and the High Energy Astrophysics Division of the Smithsonian Astrophysical Observatory;
\item Ned Wright's cosmology calculator (\citealt{Wright06}).
\end{itemize}




\bibliographystyle{mnras}
\bibliography{jetpaper-bib} 


\appendix

\section{Appendix: additional X-ray absorption}
As stated in \S~\ref{sec:xspec}, if we adopted the \citetalias{DL90} Galactic column $N_\mathrm{H}^\mathrm{Gal}=1.46\times10^{21}$~cm$^{-2}$, the deabsorbed \textit{XMM-Newton} EPIC X-ray spectra turned down towards lower energies and did not smoothly connect with the OM photometry.  
We achieved a significant improvement in the fit ($\Delta\chi^2=162$ for one additional free parameter), and a corrected shape in the deabsorbed spectra, if we allowed the $N_\mathrm{H}^\mathrm{Gal}$ to be a free parameter in our fits and increase to a value $\approx2.2\times10^{21}$~cm$^{-2}$.

We tested our other X-ray data for evidence of this additional absorprion.
We added the \textit{Swift} XRT spectra recorded between 2 August and 29 September 2015; the co-added spectrum contains 8724 counts.
A sum of two power-laws model with $\Gamma_1=2.2_{-0.1}^{+0.2}$ and $\Gamma_2=1.0_{-0.4}^{+0.3}$ has $\chi^2_\nu=357/260=1.37$ if the Galactic column is fixed to $1.46\times10^{21}$~cm$^{-2}$.
Allowing the Galactic column to be a free parameter, we find the fit improves by $\Delta\chi^2=18$ with an $F$-test probability of 99.97\%.
The Galactic column in this model is very high at $\left(4.0_{-0.8}^{+0.9}\right)\times10^{21}$~cm$^{-2}$ and the soft photon index is very steep, $\Gamma_1=4.9\pm0.7$, but clearly these parameters are poorly constrained by the limited quality of the spectrum.
We note that the shapes of the deabsorbed, co-added \textit{Swift} XRT spectra with and without the additional column agree with the corresponding \textit{XMM-Newton} EPIC spectra.
We fitted a blackbody plus power-law model fitted to the \textit{XMM-Newton} RGS spectra (taken contemporaneously with the EPIC spectra) and recorded a C-statistic 2138 with the \citetalias{DL90} value of $N_\mathrm{H}^\mathrm{Gal}$.
Increasing the Galactic column to $2.1\times10^{21}$~cm$^{-2}$ worsens the C-statistic to 2211.
However, we note that above $\approx30$~\AA\ (below $\approx0.4$ keV) the count rates in many channels are consistent with zero.
Therefore, there is not such strong evidence for a higher Galactic column in our \textit{Swift} XRT data and no evidence in the \textit{XMM-Newton} RGS spectrum.

It is unlikely that the neutral atomic hydrogen column on the line-of-sight towards 1H~0323+342 is truly this much higher than found by \citetalias{DL90}.
Whilst it is known that there are small-scale ($\sim1$--$3^{\prime\prime}$), low column density structures which may have been unseen or unresolved by H~\textsc{i} 21~cm surveys (\citealt{BenBekhti09}), we are unaware of such clumps having been detected with column densities as high as implied by our fits ($N_\mathrm{H}>10^{20}$~cm$^{-2}$).
If such a neutral absorber were in the Milky Way we would also expect to see additional reddening in our optical/UV data.
However, we find no evidence of additional redenning in our optical / UV data.  
We measured the equivalent width (EW) of the Na~\textsc{i}~D absorption line in our Keck spectrum of February 2016 to be $\mathrm{EW}=0.891$~\AA\, assuming its profile to be similar as that of the broad H$\beta$ emission line.
Using this measurement, we obtain an estimate of the extinction $A_V=0.483^{+0.098}_{-0.081}$ using the $E(B-V)$-EW(Na~\textsc{i}~D) relation of \cite{Poznanski12} and assuming the typical Milky Way $R_V=3.1$.
This value is slightly lower than the $A_V=0.706$ we derived from the literature value of the Galactic H~\textsc{i} column.

If the absorber were intrinsic to the AGN, it is possible that there was some occultation of the compact X-ray source but not the more extended optical/UV emission (e.g.\ \citealt{Risaliti07}; \citealt{Zhang17}).
We modelled the \textit{XMM-Newton} EPIC spectra with Galactic plus intrinsic columns.
For the intrinsic column we tried both neutral (\textsc{zphabs}) and partially-ionised (\textsc{zxipcf}) models.
The neutral, intrinsic column improves the fit by $\Delta\chi^2=104$ for one additional free parameter (a lesser improvement than the additional Galactic column) and we find $N_\mathrm{H}^\mathrm{int}=(8\pm1)\times10^{20}$~cm$^{-2}$.
With the \textsc{zxipcf} model we obtain a very low $\xi$ value, indicating weakly ionised material, and a high column $N^\mathrm{int}_\mathrm{H}=(2\pm1)\times10^{21}$~cm$^{-2}$.  
The improvement in the fit is only $\Delta\chi^2=11$ for three additional free parameters and this additional ionised intrinsic absorber did not correct the shape of the deabsorbed spectra.

Despite extensive modelling, we have been unable to find a physically plausible model with a column density fixed at the \citetalias{DL90} value which both reduces the residuals and also gives a corrected shape of the deabsorbed soft spectrum that fits the UV data.
We adopted the increased neutral Galactic column as being the simplest model solution which improved our fits and the shape on the intrinsic spectrum.
 
As we showed in \S~\ref{sec:optxconv}, this allows us to fit an energy-conserving accretion disc model which reproduces the optical/UV to hard X-ray data and returns parameters typical of a NLS1.
Our jet models would not be substantially changed if we had proceeded with an X-ray spectrum deabsorbed through the \citetalias{DL90} column density value.

\bsp	
\label{lastpage}
\end{document}